\PassOptionsToPackage{dvipsnames}{xcolor}	
\documentclass[12pt,letter]{article}

\usepackage[T1]{fontenc}
\usepackage{lmodern}
\usepackage{setspace}
\usepackage{etoolbox}
\makeatletter
\patchcmd{\appendix}{\@Alph}{\@Roman}{}{}
\makeatother

\usepackage{amsmath}
\usepackage{amssymb}
\usepackage{amsthm}
\usepackage{mathtools}
\usepackage{mathrsfs}
\usepackage{breqn}
\usepackage{bigints}

\usepackage[margin=1 in]{geometry}
\usepackage{enumerate}
\usepackage{enumitem} 
\usepackage{xcolor}
\usepackage[hidelinks]{hyperref}
\usepackage{float}
\usepackage{indentfirst}
\usepackage{caption}
\usepackage{fancyhdr}
\usepackage{tcolorbox}
\usepackage{multirow,array}

\usepackage{tikz}
\usetikzlibrary{decorations.pathreplacing}
\usepackage{mathtools}
\usetikzlibrary{positioning}
\usepackage{graphicx}

\newcommand{\mcal}{\mathcal}

\renewcommand{\epsilon}{\varepsilon}
\newcommand{\E}[2][]{\mathbb{E}_{#1}\left[#2\right]}

\newcommand{\blue}[1]{\color{blue}#1 \color{black}}

\newcommand\independent{\protect\mathpalette{\protect\independenT}{\perp}}
\def\independenT#1#2{\mathrel{\rlap{$#1#2$}\mkern2mu{#1#2}}}

\newtheorem{theorem}{Theorem}
\newtheorem{lemma}{Lemma}
\newtheorem{proposition}{Proposition}

\newtheorem{claim}{Claim}

\newtheorem*{statement*}{Result}

\theoremstyle{definition}
\newtheorem{definition}{Definition}

\DeclareMathOperator*{\argmax}{\arg\!\max}
\DeclareMathOperator*{\argmin}{\arg\!\min}
\DeclareTextFontCommand{\emph}{\slshape}

\usepackage{natbib}
\nocite{*}

\title{Simple Paired Combinatorial Assignment}

\author{Eric Gao\thanks{Department of Economics, Massachusetts Institute of Technology. ericgao@mit.edu.} 
\footnote{I am especially grateful to Paul Milgrom for valuable advice throughout this project. I would like to thank Jose Blanchet, Matthew O. Jackson, Daniel Kornbluth, Tina Li, Irene Lo, Daniel Luo, Christopher Makler, conference participants at Georgetown, and seminar participants at Stanford for helpful feedback and discussions.}}

\date{\today}

\begin{document}
	
\maketitle
	
\begin{abstract}
    Consider a university assigning students to courses and dorms. While many mechanisms are available, they each have their own drawbacks. Running serial dictatorship once for all goods is highly unfair, but running serial dictatorship separately for each matching problem is inefficient---Pareto improvements can be found via students jointly trading their allocated course and dorm. Alternatively, competitive equilibrium from equal incomes scales combinatorially in the number of items, making implementation and preference elicitation difficult. This paper considers \textit{paired serial dictatorship}: a novel mechanism where agents signal relative preferences that determine their priority in each market. Any deterministic allocation that arises in equilibrium is Pareto efficient and envy-free, highlighting how seemingly innocuous tie-breaking is the key barrier to optimality and fairness. When agents differ only in relative preferences, paired serial dictatorship ex-ante Pareto dominates running random serial dictatorship independently in each market. Such gains exist even when agents behave simplistically.
\end{abstract}

\noindent \textbf{Keywords}: Combinatorial Allocation, Serial Dictatorship, Course Allocation. \\
\noindent \textbf{JEL Codes}: D47.
	
\newpage
\onehalfspacing
\section{Introduction}

Combinatorial allocation problems are canonically difficult, especially when preferences exhibit complementarities and transfers cannot be used. While a wide range of potential mechanisms have been developed to solve this problem, each possesses its own benefits and drawbacks. One class of mechanisms involves adaptations of competitive equilibrium from equal incomes (CEEI) where each agent is endowed with some fictitious currency, reports valuations over bundles, and a competitive equilibrium of the resulting economy is found and implemented. Unfortunately, this approach faces two shortcomings. First is the non-existence of competitive equilibrium in the presence of complements, which has required modifications of the mechanism to settle for approximate competitive equilibrium from approximately equal incomes (see \cite{budish_2011_the}).\footnote{\cite{nguyen_2024_nearsubstitute} considers different approximations under alternative restrictions on preferences.} The other major drawback is that agents need to report a \textit{cardinal} utility for every potential \textit{bundle} of goods to run CEEI, a problem that grows combinatorially in the number of goods. As such, the mechanism has only been used to allocate courses at small schools with limits on what preferences students can submit.\footnote{In particular, students are only allowed to report a cardinal utility for each course and a (potentially negative) modifier for pairs of courses to indicate that they are either substitutes or complements. Students cannot report preferences over bundles of size three or greater \citep{budish_2012_the}.} However, the mechanism is able to deliver (approximately) fair and efficient outcomes.

Another commonly used mechanism is random serial dictatorship (RSD), where agents are ordered in some way and choose their favorite bundle in sequence from what is still available. While this mechanism is Pareto efficient and simple for agents to engage with, it is highly ex-post unfair: Agents with a higher priority get to choose \textit{all} their goods before agents with a lower priority get to choose \textit{any} of their goods. On the other hand, if multiple instances of RSD were used to allocate different classes of goods in an attempt to restore fairness, ex-post Pareto efficiency is lost as running things independently (which will be called independent RSD) ignores agents' relative preferences between different classes of goods.

Are there any mechanisms that achieve both the fairness of CEEI and the simplicity of serial dictatorship (while still guaranteeing efficiency)? This paper introduces and evaluates \textit{paired serial dictatorship}---a novel mechanism for combinatorial allocation problems that can be decomposed into two components. In the first component, it must be that agents have single-unit demand and do not exhibit complementarities or substitutabilities with the remainder of the allocation problem. In the other component, arbitrary combinatorial preferences over bundles of any size are allowed. Paired serial dictatorship operates by first asking agents to report their relative preferences between the two components (e.g. ``On a scale of one to ten, how much do you care about your courses versus your dorm?'' in the context of course and dorm assignment) and uses those reports to order agents and run serial dictatorship in either market. Agents that report a higher relative preference in one component choose earlier in that component but later in the other component. This paper is the first to analyze \textit{how} agents should be ordered under serial dictatorship, in the case where there are two classes of goods to be allocated.\footnote{Additionally, paired serial dictatorship is a minimal change over independent RSD. Current systems that run serial dictatorship for courses and dorms stay the same; the only difference is central school administration simply telling departments what orderings to use instead of them drawing priorities randomly.}

Our restriction of having one dimension of preferences being separable can arise by either making assumptions within a given allocation problem or by expanding the scope of the problem itself. This property also arises naturally in many of the canonical examples (taken from \cite{budish_2011_the}) of combinatorial allocation without transfers: 
\begin{itemize}
    \item In course allocation, certain fun or extracurricular courses (\href{https://explorecourses.stanford.edu/search?view=catalog&filter-coursestatus-Active=on&q=DANCE%2046:%20Social%20Dance%20I&academicYear=20192020}{\blue{social dance}at Stanford}, \href{https://english.fas.harvard.edu/english-183ts-taylor-swift-and-her-world}{\blue{Taylor Swift}at Harvard}, \href{https://scl.cornell.edu/coe/ctc/tree-climbing-courses}{\blue{tree climbing} at Cornell}) are often orthogonal to a student's primary coursework. Alternatively, universities may consider jointly allocating courses and dorms (which are often independently allocated via random serial dictatorship).
    \item In worker scheduling, weekly tasks can be paired with the allocation of parking spaces or who has priority to take paid time off around holidays.
    \item In the allocation of players to sports teams, goalies (or other positions) may exhibit less complementarities or substitutabilities with the rest of the team compared to other players. Alternatively, sports governing bodies may jointly allocate players and convenient travel schedules.\footnote{Major league soccer (MLS) organizes a ``SuperDraft'' every year that runs serial dictatorship for new players that just graduated college. Teams are ordered by season performance, but ties are broken randomly among teams that performed the same. Furthermore, in 2018 the team that traveled the most traveled more than \href{https://www.mlssoccer.com/news/which-team-will-have-travel-most-least-2018}{\blue{twice}}the distance of the team that traveled the least.}
\end{itemize}
To fix ideas, we will use the application of allocating courses and dorms to students for the remainder of the paper: agents will be called students, the class of goods where agents have single-unit demand and are evaluated independently will be dorms, and the remainder of the market where combinatorial preferences are permitted will be courses.

Paired serial dictatorship induces the desired screening: Students who care more about courses (or dorms) do indeed report that, regardless of other students' reports and preferences. This separation, in turn, leads to paired serial dictatorship allocating better courses to students that care more about courses and better dorms to students that care more about dorms. However, multiple students will generally report the same relative preference, which means that tie-breaking is still required at times. Whenever this tie-breaking does not influence the resulting allocation---so signaling and redistribution of priorities fully resolves coordination frictions among students---the resulting deterministic allocation is Pareto efficient and ex-post envy-free (Theorem \ref{deterministic_pareto}). As such, a significantly simpler mechanism is able to deliver the promises of CEEI without ever requiring students to report complete cardinal preferences over all potential bundles.\footnote{This result is similar in spirit to the analysis done in \cite{feigenbaum_2020_dynamic}, which considers a two-stage matching process for students to schools, incorporating preference updating while on waitlists. Under their mechanism, lotteries are reversed between the initial and second rounds of matching. Their key technical assumption is an ``order condition'' which guarantees that if one school is more popular than another in the initial round, it remains more popular in the second round. This assumption is imposed to guarantee that the time each school reach capacity is well-behaved. In our model, the assumption of a deterministic allocation plays the same role, ensuring that the time each course or dorm reaches capacity corresponds precisely to the mass of students revealing a certain signal.} While any student taking up a spot in a course or dorm has an externality on every other student, each student's ability to optimize the signal they choose to reveal in equilibrium outweighs those externalities. As such, the need for ad-hoc tie-breaking when student reports are not sufficiently nuanced is the key driver of inefficiency as the mechanism simply does not have the necessary information to align priorities with preferences. In the worst case, full randomization under independent RSD is a key driver of suboptimality.

The remainder of the paper is organized as follows. We first work through a simple example demonstrating the shortcomings of independent RSD and how paired serial dictatorship resolves inefficiencies. Next, Section \ref{lit_review} reviews related literature and discusses potential drawbacks of alternative course/combinatorial allocation mechanisms. Section \ref{model} develops the model and provides two comparative statics about student behavior under paired serial dictatorship, while Section \ref{welfare} analyzes welfare under paired serial dictatorship. In Section \ref{simulations}, an economy of students choosing courses and dorms is simulated to examine the performance of paired serial dictatorship when the assumptions needed for theoretical guarantees do not hold. Finally, Section \ref{conclusion} concludes. \ref{Appendix A} collects notation used throughout the appendices, proves that paired serial dictatorship is well-defined, and shows that a pure-strategy Nash equilibrium exists. All proofs are collected in \ref{Appendix B}.

\subsection{A Motivating Example}

We begin with an example illustrating the potential issues with independent RSD.  Suppose there are two students, Chris ($C$) and Doug ($D$), two courses $c_1$ and $c_0$, and two dorms $d_1$ and $d_0$. Each student demands one course and one dorm, and each course or dorm has a capacity of one. Both students prefer $c_1 \succ c_0$ and $d_1 \succ d_0$ (i.e. subscripts denote quality), but Chris cares more about courses and Doug cares more about dorms. To parameterize this, let
$$U_C(c_\ell,d_m) = 10\ell + m; U_D(c_\ell,d_m) = \ell + 10m.$$
Under independent RSD, there are four combinations of which student chooses when in either market, each with equal probability:
\begin{enumerate}
    \item If Chris chooses first in both markets, they receive $c_1, d_1$ and a utility of $11$ while Doug receives $c_0, d_0$ and a utility of $0$.
    \item If Chris chooses courses first and dorms second, they receive $c_1, d_0$ and a utility of $10$ while Doug receives $c_0, d_1$ and a utility of $10$.
    \item If Chris chooses courses second and dorms first, they receive $c_0, d_1$ and a utility of $1$ while Doug receives $c_1, d_0$ and a utility of $1$.
    \item If Chris chooses second in both markets, they receive $c_0, d_0$ and a utility of $0$ while Doug receives $c_1, d_1$ and a utility of $11$.
\end{enumerate}
Outcomes (1) and (4) are Pareto efficient yet highly unfair and do not maximize total surplus; these are also the outcomes selected by running serial dictatorship once and letting students pick both their course and dorm. Outcome (3) is Pareto efficient \textit{within} each market if looking at the market for courses or dorms separately; however it is Pareto inefficient in the overall economy as it is Pareto dominated by outcome (2). Additionally, each student envies the other student's bundle and both would be better off by trading. Finally, Outcome (2) is truly Pareto efficient. In this case, random tie-breaking just happens to align with individual preferences: the student who cares more about courses just happens to choose courses first while the student who cares more about dorms just happens to choose dorms first. 

What happens if paired serial dictatorship is used? Suppose the designer simply asked students if they care more about courses or dorms. If both students give the same answer, we return to uniform randomization leading to each student receiving an (expected) utility of $(11+10+1+0)/4 = 5.5$. Otherwise, each student picks first in the market they say they care more about. As such, this mechanism induces the following normal form game:

\begin{figure}[h]
    \centering
    \setlength{\extrarowheight}{2pt}
    \begin{tabular}{cc|c|c|}
      & \multicolumn{1}{c}{} & \multicolumn{2}{c}{Doug}\\
      & \multicolumn{1}{c}{} & \multicolumn{1}{c}{$C$}  & \multicolumn{1}{c}{$D$} \\\cline{3-4}
      \multirow{2}*{Chris}  & $C$ & $5.5,5.5$ & $10,10$ \\\cline{3-4}
      & $D$ & $1,1$ & $5.5,5.5$ \\\cline{3-4}
    \end{tabular}
    \caption{Normal form game induced by a paired serial dictatorship mechanism.}
\end{figure}

Chris is the row player, Doug is the column player, and $x, y$ denotes a payoff of $x$ to Chris and a payoff of $y$ to Doug. The unique Nash equilibrium of this game is for Chris to reveal $C$ and for Doug to reveal $D$; in fact, it is a strictly dominant strategy for students to ``truthfully'' report their relative preferences over courses and dorms. 

Furthermore, this result holds for arbitrary cardinal utilities of Chris and Doug (as long as Chris cares more about courses and Doug cares more about dorms); the mechanism arrives at the efficient outcome without ever knowing precise cardinal values. Unfortunately, the trade-off here is that a strictly dominant strategy may not always exist for more general economies. As such, paired serial dictatorship requires students to engage in equilibrium reasoning in exchange for simplicity when reporting preferences. In the course and dorm allocation problem (and many of the other problems discussed above), the allocation mechanism is run year after year and there is much discussion about what courses and dorms are in high demand. Students often talk about their draw times and what was available by the time they got to choose. Furthermore, students need not fully compute best responses to others’ underlying utility functions, but rather only need to compute expectations over when certain courses and dorms run out. Additionally, it is not the end of the world for any student even if they fail to behave optimally: Any courses and dorm that a student gets under paired serial dictatorship can also arise from independent RSD. Finally, simulations (in Section 5) suggest that the cost of bounded rationality is negligible.

\section{Related Literature}\label{lit_review}

This paper most directly contributes to the literature on competitive equilibrium from equal incomes and its applications to course allocation. Contrary to other examples of one-sided matching or mechanism design without transfers, course allocation imposes additional structure: Students have preferences over bundles of courses that may exhibit both complementarities or substitutabilities while scheduling or balance constraints may rule out certain bundles altogether. The literature on course allocation, starting with \cite{budish_2010_finding} and \cite{budish_2011_the}, has focused on the (approximately) competitive equilibrium from equal incomes (CEEI) mechanism, where students are each endowed with (approximately) the same amount of some fake currency, submit cardinal preferences over bundles of courses, and receive the market-clearing allocation. The CEEI mechanism has several attractive theoretical properties: Outcomes are approximately efficient, the mechanism is strategyproof when the number of students participating in the market is large, and each student gets at least their maximin share of utility and only ever envies another student's bundle of courses up to a single course. \cite{budish_2017_course} details implementing the approximate CEEI mechanism at Wharton Business School where 1,700 students are assigned to 350 courses. Outcomes in practice are fair and efficient, reflected in student surveys where respondents report higher satisfaction and gains in perceived fairness. More recently, \cite{kornbluth_2021_undergraduate} incorporates priorities (for example, students with seniority are favored over newer students) into the CEEI mechanism to form a new mechanism called Pseudo-Market with Priorities (PMP). They then use data from around 6,000 students across 7 colleges selecting from 756 courses to calibrate student utilities and evaluate PMP against deferred acceptance and serial dictatorship, clearing each college separately. PMP generally performs the best, leading to higher average utilities and lower standard deviations in utility with respect to random initializations. Unfortunately, such market-based mechanisms are a poor fit to addressing undergraduate course allocation writ large. In \cite{kornbluth_2021_undergraduate}, each college had no more than 1,642 students or 269 courses, but many other institutions have an order of magnitude more students or courses. It would be infeasible to ask students to provide cardinal utilities for every course they might be interested in, and solving for the market-clearing allocation would be difficult. Furthermore, joint trades of courses and dorms can still be found even when courses are allocated via CEEI. 

Instead, \cite{budish_2012_the} considers the Harvard Business School ``Draft'' mechanism, where students are first assigned some order, choose their first class in order, choose their next class in reverse order, and so on. There are some similarities between student behavior under the Draft mechanism and paired serial dictatorship, formalized in the discussion following Proposition \ref{competition_seperation}. \cite{bichler_2021_randomized} investigates how randomized scheduling mechanisms compare to first-come-first-serve mechanisms. In particular, they find that the probabilistic serial mechanism generally outperforms random serial dictatorship when both are adapted to handle bundles of objects, but differences are small. Unfortunately, bundled probabilistic serial is not strategyproof, which students take advantage of in experiments using the mechanism. \cite{bhalgat_2011_social} carries out a similar comparison between probabilistic serial and random serial dictatorship, but for the case of one-to-one matching. \cite{romeromedina_2024_strategic} evaluates deferred acceptance, immediate acceptance, and conditional acceptance algorithms in the context of course allocation.

Beyond the domain-specific literature of course assignment, we also relate to the literature identifying the limits of what (random) serial dictatorship mechanisms can achieve. Much work has gone into axiomatizing such mechanisms: \cite{svensson_1999_strategyproof} demonstrates that a mechanism is deterministically non-bossy, strategyproof, and neutral if and only if it is serial dictatorship. \cite{bade_2020_random} considers random mechanisms and shows that random assignment of agents into roles in any mechanism that is ex-post Pareto efficient, non-bossy, and strategyproof leads to the same distribution over outcomes as random serial dictatorship while \cite{basteck_2024_an} shows that random serial dictatorship is the only mechanism that satisfies symmetry, ex-post Pareto efficiency and probabilistic (Maskin) monotonicity. \cite{imamura_2024_efficient} utilizes such characterizations to show that an outcome is Pareto efficient only if it can be found via some serial dictatorship mechanism. Finally, \cite{kikuchi_2024_a} takes an ex-ante approach, showing that ex-ante efficiency and Bayesian incentive compatibility can only be achieved by dictatorial mechanisms. 

The other direction prior work has considered is extensions of the Shapley–Scarf economy to allow for multiple types of goods. Unfortunately, most results have been negative. \cite{konishi_2001_on} finds that even when preferences are strict and separable, there may not be any allocations in the (strict) core; if non-empty, the core may not be unique; and the set of allocations in the core may differ from the set of allocations that arise in competitive equilibrium. ``Furthermore, there is no Pareto efficient, individually rational, and strategy-proof social choice rule.'' \cite{ppai_2001_strategyproof} analyzes what allocation rules are strategyproof, non-bossy (and satisfy citizen sovereignty, which states that for every possible allocation, there is some profile of preferences that leads to it). Only sequential dictatorships---allocation rules similar to serial dictatorships, but the order agents choose is endogenously determined by their reports---satisfy these three conditions. To get around the tension between Pareto efficiency and strategyproofness, \cite{klaus_2007_the} considers \textit{second-best incentive compatible}---a mechanism is second-best incentive compatible if no other incentive compatible mechanism Pareto dominates it---and shows that finding core allocations in the market for each type of good (independent of markets for other types of goods) is second-best incentive compatible. \cite{anno_2016_on} generalizes this result, finding that using strategyproof and non-wasteful allocation rules market-by-market results in a second-best incentive compatible allocation rule in the aggregate. We contribute to this literature by considering what can be attained when strategyproofness is dropped. In particular, the universe of reports available to each student is much smaller than the set of possible student types (a potential desiderata for allocation problems when reporting entire types is infeasible), which in and of itself leads to traditional notions of strategyproofness to break down.

Finally, we relate to the literature on mechanism design without transfers. Under a standard mechanism design framework, agent utilities are assumed to be quasilinear in money, leaving the designer with great flexibility to alter incentives by changing payments associated with different outcomes. Without transfers and a one-dimensional setting, we end up back at the one-sided matching world. When there are two dimensions like the setting of joint course and dorm assignment, the designer has more flexibility to use goods in one market as a shadow price on goods in the other market. Taking advantage of this, \cite{kesten_2021_strategyproof} studies mediation in cases where two mediators have diametrically opposed ordinal preferences over two separate issues. A strategyproof and efficient mechanism exists only if every outcome in one dimension can be compensated by a sufficiently large concession in the other dimension, giving the designer sufficient flexibility to alter incentives. However, the literature on mechanism design without transfers has generally focused on the case where there are a large number of issues. The early paper of \cite{jackson_2007_overcoming} notes that any ex-ante efficient distribution over outcomes is attainable when enough copies of a decision are linked with itself. \cite{ball_2023_quota} works in the same setting but considers finite independent copies of a social choice problem and analyzes the performance of quota mechanisms, mechanisms which fix the distribution over outcomes across decisions. \cite{horner_2015_dynamic} and \cite{balseiro_2019_multiagent} consider dynamic mechanism design without transfers in an infinite-horizon case and find that optimal mechanisms take the form of giving each agent a ``utility promise'' in the future which acts like a transfer. When there are enough periods, the designer can use design utility promises to act as transfers in the current period. \cite{ezzatelokda_2023_a} studies a similar problem but proposes a ``karma'' mechanism: Instead of utilizing utility promises, each agent has some amount of a fake currency called ``karma'' that they can either spend in the current period or save and use in the future. The mechanism considered in this paper has similar characteristics: The signal a student sends to the mechanism about how much they care about their courses can be thought of as how much karma they're willing to spend on an earlier course enrollment time, while any leftover karma goes towards obtaining an earlier dorm selection time.

\section{Model}\label{model}

There is a continuum of students denoted by the interval $\mcal I = [0,1]$ and endowed with the Lebesgue measure $\mu$ to model a large economy where each individual has zero aggregate effect. Let $i \in \mcal I$ denote an individual student. There is a finite set $\mcal C$ of courses and a finite set $\mcal D$ of dorms (when thinking about other applications, $\mcal C$ and $\mcal D$ can be interpreted as arbitrary goods). Each $c \in \mcal C, d \in \mcal D$ denotes an individual course or dorm, while $C \subset \mcal C, D \subset \mcal D$ denotes a set of courses or dorms. Let $q: \mcal C \cup \mcal D \to \mathbb{R}$  denote a mapping from courses and dorms to how many students can enroll in a course or live in a dorm (so $q(c_1)$ represents the mass of students that may be enrolled in course $c_1$).
Student $i$ receives utility from a bundle of courses $C$ and a dorm $d$ according to
$$u_i(C,d) = v_i(C) + w_i(d)$$
so preferences are additively separable across courses and dorms. This assumption is meaningful beyond analytic tractability. At many institutions, course and dorm selection happen at different points in time: For example, undergraduates who are guaranteed housing need to be matched early on before graduate students can select their rooms, whereas course enrollment cannot happen until professors figure out what and when they are teaching. As such, institutional constraints may restrict students from reporting richer preferences over courses and dorms. On the student's side, preferences over one market may include \textit{expectations} of the other market. One natural source of complementarities comes from students wanting to live close to their classes. In this case, even if students do not know which classes are available (both in general or for them to enroll in), they still know that large lectures will be held in large lecture halls and classes in certain departments will be near the department building. Thus, a statistics student may have ``distance to the statistics building'' as part of their utility from dorms without violating additive separability.

Each student has access to some finite and ordered set of signals $\mcal S = \{1,...,S\}$ for some $S \in \mathbb{N}$. For example, simply asking students if they care more about courses or dorms is equivalent to setting $\mcal S = \{1, 2\}$; school administration may ask students to rank their relative preference between courses and dorms on a scale of one to ten, in which case $\mcal S = \{1, 2, ..., 10\}$. Let $\hat{s}(i)$ denote the signal that student $i$ reports. To account for ties, let $r^c, r^d: \mcal I \to [0,1]$ be two measurable and bijective functions (we will call them permutations of the unit interval). Let $R$ be a probability distribution over the space of all permutations of the unit interval from which $r^c, r^d$ are drawn.\footnote{For a discussion on the existence of distributions over the set of all bijective functions from the unit interval from itself (such as the uniform distribution), see footnote 9 of \cite{takahashi2010community}.} As regularity conditions, suppose:
\begin{enumerate}
    \item[(A1)] Measurability of Preferences: For every $C, d$, the maps $v_{(\cdot)}(C): \mcal I \to \mathbb{R}$ and $w_{(\cdot)}(d): \mcal I \to \mathbb{R}$ are measurable.
    \item[(A2)] Symmetric Tie-breaking: If $B$ is a set of permutations of the unit interval that is $R$-measurable, then for all $i, j \in \mcal I$,
    $$R(B) = R(B \circ \sigma_{ij}) = R\left(\{(r^c, r^d): (r^c \circ \sigma_{ij}, r^d \circ \sigma_{ij}) \in B\}\right)$$
    where $\sigma_{ij}$ is the permutation of $i$ and $j$.
\end{enumerate}
The \textit{paired serial dictatorship} mechanism proceeds as follows:
\begin{enumerate}
    \item The designer chooses $\mcal S$.
    \item Students observe the mechanism and simultaneously choose signals $\hat{s}(i)$.
    \item Nature draws $r^c, r^d$.
    \item Run serial dictatorship for bundles of courses ordering students via $\hat{s}(i) + r^c(i)$.
    \item Run serial dictatorship for dorms ordering students via $-\hat{s}(i) + r^d(i)$
    \item Students receive the courses and dorms prescribed by Steps (4) and (5).
\end{enumerate}

Standard serial dictatorship reasoning gives that each student selects their favorite set of courses/dorms out of all courses/dorms that are available at the time of their choice. As such, given $\mcal S$, each student effectively only chooses a signal to send, which is aggregated into the function $\hat{s}$.\footnote{Another way to think about signals is as a fictitious currency (as seen in \cite{budish_2017_course} applied to course allocation and \cite{prendergast_2022_the} applied to food allocation). Under this interpretation, each student has initial income $|S|$ and can allocate integer amounts to bid on an earlier course or dorm selection time.} Additionally, observe that independent RSD is a special case of paired serial dictatorship when $\mcal S$ is taken to be a singleton. In a more general sense, paired serial dictatorship is ``worst-case'' bounded by random serial dictatorship: Any outcome of paired serial dictatorship is possible under random serial dictatorship.

\ref{Appendix A} shows that the game and the resulting allocation are well-defined, and additionally proves existence of a pure strategy Nash equilibrium.\footnote{Unfortunately the equilibrium may not be unique as the same allocation may be induced using different sets of signals. It is an open question as to whether or not all equilibria are payoff-equivalent.} The proof utilizes \cite{schmeidler_1973_equilibrium}'s existence results. However, this model does not map perfectly into their setting as a student's payoff from playing a given strategy is not necessarily continuous in other students' strategies: For students revealing a signal on the boundary between two signals needed to get a certain course or dorm, their utility may jump when others' strategies are perturbed. Fortunately, these discontinuities are well-behaved; there are finitely many such discontinuities (at most one for every course or dorm) and they are all downward jumps.

Before proceeding to discussing welfare, we analyze two forces---relative preferences between courses and dorms and competition for certain courses or dorms---that influence student behavior. First, students do indeed separate themselves based on their relative preferences between courses and dorms: A student who cares relatively more about the same courses compared to some other student will report a higher value of $s$. 

\begin{definition}
    Student $i$ \textbf{cares more (less) about the same courses} as student $j$ if:
	\begin{enumerate}
		\item The two students have the same preference ordering over bundles of courses: $v_i(C) \geq v_i(C')$ if and only if $v_j(C) \geq v_j(C')$;
		\item $v_i(C) - v_j(C)$ is increasing (decreasing) with respect to the common order over courses;
		\item Dorm preferences are identical: for all dorms, $w_i(d) = w_j(d)$.
	\end{enumerate}
\end{definition}

Analogous definitions hold for dorms. Functionally, a student caring more (less) about the same courses is equivalent to that student caring less (more) about the same dorms.

\begin{proposition}[Separation via Relative Preferences]\label{preferences_seperation}
    In any equilibrium, students who care more about the same courses (dorms) will report higher (lower) signals.
\end{proposition}

Second, competition between students for the same courses or dorms also influences the signal each student reports. In particular, courses that face higher demand require students to report higher signals to attain an enrollment time that is early enough. On the other hand, if there ever is slackness in a student's reported signal in the sense that reporting a lower signal does not lead to receiving worse courses, then a student would benefit from reporting that lower signal to take advantage of an earlier dorm selection time without compromising the courses they receive. As such, students who prefer less competitive courses report lower signals, while students who prefer less competitive dorms report higher signals. 

\begin{proposition}[Separation via Competitiveness]\label{competition_seperation}
    Fix any $\hat{s}$. If student $i$ receives the same distribution of courses (dorms) when sending $s'$ instead of $\hat{s}(i)$ and $s'$ is lower (higher) than $\hat{s}(i)$, then student $i$ is weakly better off reporting $s'$ than $\hat{s}(i)$.
\end{proposition}

The intuition behind this result is similar to the intuition behind Theorem 1 of \cite{budish_2012_the} which analyzes manipulating the Draft mechanism by listing a highly demanded but less preferred course ahead of a less demanded but more preferred one. However, the precise mechanism differs: In \cite{budish_2012_the}, the order of selection is fixed, and such deviations allow a student to secure a competitive class before worrying about securing a non-competitive class. Under paired serial dictatorship, the deviation changes the \textit{order of choice itself}: Student $i$ chooses later relative to other students in the market for some non-competitive good, but is able to choose earlier in the other market. As such, the current approach is structurally different from an alternative approach where dorms are simply interpreted as another course to be allocated under the Draft mechanism.

\section{Welfare}\label{welfare}

Our first welfare result states that when students separate themselves via relative preferences, that separation does indeed translate into envy-freeness between those students. This holds regardless of tie-breaking. Unfortunately, this does not hold for students that ultimately reveal the same signal: random tie-breaking means that if one student cares more/less about the same courses/dorms than another student, it could still be that those students would be better off trading their bundles.

\begin{proposition}[No Ex-Post Swaps]\label{seperation_no_swap}
    Let $\hat{s}$ be an equilibrium. If student $i$ cares more about the same courses as student $j$ and $\hat{s}(i) \neq \hat{s}(j)$, then for all $r^c, r^d$, students $i$ and $j$ would \textit{never} both prefer each others' courses and dorms.
\end{proposition}

In general, no student will ever prefer the \textit{distribution} over courses and dorms any other student receives to the distribution over courses and dorms that they receive. If student $i$ ever ex-ante envied student $j$, student $i$ could have always revealed the same signal as student $j$ and picked the same courses or dorm when it is their turn to choose (as tie-breaking is symmetric).


\begin{proposition}[Ex-Ante Envy-Freeness]\label{FOSD_envy_free}
    Any allocation that arises in equilibrium is envy-free: No student would ever prefer any other student's distribution over courses and dorms to their own distribution over courses and dorms.
\end{proposition}

As Proposition \ref{FOSD_envy_free} holds for any paired serial dictatorship mechanism, a corollary is that even independent RSD\footnote{Recall that independent RSD can be recovered from paired serial dictatorship if the signal space is a singleton.} produces a result that is ex-ante envy-free. However, ex-ante envy-freeness is a weak property that does not speak to overall efficiency. It turns out that randomization, necessitated by tie-breaking, is the key barrier to establishing results about efficiency (and ex-post envy-freeness).

\subsection{Deterministic Efficiency and Fairness}

Students' choices of signals are determined by their relative preferences between courses and dorms (Proposition \ref{preferences_seperation}) and how competitive their favorite courses and dorms are (Proposition \ref{competition_seperation}). However, the final bundle a student receives is also mediated by tie-breaking, which could either harm or help coordination as seen in the motivating example. Without signaling, the mechanism can only resort to ad-hoc (random) tie-breaking which is uncorrelated with relative preferences. On the other hand, when tie-breaking does not play a role and paired serial dictatorship induces a deterministic allocation (the courses and dorm each student is assigned to are constant in $r^c, r^d$), the resulting outcome is Pareto efficient. In this case, allocating priorities via paired serial dictatorship is sufficient to resolve students' conflicts of interest in the course and dorm assignment stage.

\begin{theorem}\label{deterministic_pareto}
    Any equilibrium of paired serial dictatorship that induces a deterministic allocation is Pareto efficient and envy-free.
\end{theorem}

The proof relies on showing that if some student were strictly better off under an alternative allocation that makes no students worse off, then that student must have had a profitable deviation from the original equilibrium. As such, even though we might expect externalities to arise from students switching priorities (one student reporting a higher signal to secure a spot in a certain class leads to some other student no longer being able to register for that class), each student best-responding to $\hat{s}$ in equilibrium fully mediates these externalities. At deterministic allocations, the ex-ante distribution over allocations is degenerate, and Proposition \ref{FOSD_envy_free} directly implies that deterministic allocations are envy-free.

This result relies heavily on the induced allocation being deterministic. At random allocations, alternative definitions of Pareto efficiency must be used; the tie-breaking mechanism is instrumental for student utilities; and the link between signals and the courses and dorms attainable at those signals breaks down. Unfortunately, equilibria that induce deterministic allocations are not guaranteed to exist. If the signal space is too small, students are constrained in how granular they can report relative preferences; in the worst case of $\mcal S$ being a singleton, we arrive back at the complete randomization of independent RSD. On the other hand, expanding $\mcal S$ increases the expressivity of each student's signal, crowding out tie-breaking from being the key determinant of who gets what. Furthermore, this can only weakly increase the set of equilibria that induce deterministic allocations. 

\begin{proposition}\label{signal_expansion}
    Fix students, courses, and dorms. If $\hat{s}$ is an equilibrium that induces a deterministic allocation when the signal space is $\mcal S$, then for any $\mcal S' \supset \mcal S$, there exists an equilibrium $\hat{s}'$ that induces the same deterministic allocation.
\end{proposition}

Beyond insufficient signal spaces, student preferences may make it so that tie-breaking is required at \textit{every} equilibrium. For example, suppose there was some dorm $d$ with capacity $q(d)$, but
$$w_i(d) > \left(1+\frac{1}{q(d)}\right) \left[\max_{i, C', d' \neq d} \{v_i(C') + w_i(d')\} - \min_{i, C', d' \neq d} \{v_i(C') + w_i(d')\}\right] \text{ for all } i$$
so each student values $d$ more than $1+1/q(d)$ times more than their greatest difference in utility between any two bundles without $d$. Then, each student gets dorm $d$ with probability (at least) $q(d)$ if they report the lowest signal; that probability to get dorm $d$ outweighs all other considerations. As such, the unique Nash equilibrium (for any signal space) is for all students to report the lowest signal, leading back to full randomization. Hyper-competition for dorms is a practical concern that many students share. At worst, every student reporting the same signal leads back to independent randomization, while if a few students report something else, those students are better off under paired serial dictatorship than under independent RSD. This ``free'' improvement over independent RSD we get is a consequence of our large-market approach. In large, finite economies, this corresponds to the intuition that any one individual's deviation has a vanishingly small probability of impacting any other student's bundle.

A potential design change to get around this issue would be to draw $r^c, r^d$ first, tell students their tie-break number, and then ask students to report signals. In the above example, only students $i$ with $r^d(i) \geq 1-q(d)$ have a chance to get dorm $d$ while the remaining students now know not to embark on the lost cause of reporting the lowest signal to get dorm $d$. However, this approach still runs into the issue of randomization being instrumental in whether or not the resulting allocation is efficient - the mass of students who receive the highest $r^d$ may not be the students that value dorm $d$ the most relative to other potential bundles. Randomizing before students report signals and expansions of the signal space are evaluated on simulations in Section 5.

\subsection{Homogeneous Preferences}

Without a deterministic equilibrium, paired serial dictatorship may still be a Pareto improvement over independent RSD, but sharp characterizations become more difficult. In general, students' signals depend on the curvature of their induced utility from courses or dorms as a function of $s$. With fully arbitrary preferences, curvature for different students can vary due to their own utility function but also the competitiveness of certain courses or dorms. One way to get around this issue is to assume some degree of homogeneity in student preferences: Suppose we ignore the role that competition for certain highly demanded courses or dorms plays, and instead focus on relative preferences. To do so, suppose that students have the same ordering over sets of courses or dorms and differ only in their relative preferences. In this special case, paired serial dictatorship is an ex-ante improvement over independent RSD.

\begin{proposition}[Ex-Ante Improvement]\label{ex_ante_improvement_1}
    Suppose students had common values for courses and dorms and only differed in their relative preferences, so
    $$u_i(C, d) = \frac{\lambda_i}{1+\lambda_i}v(C) + \frac{1}{1+\lambda_i}w(d).$$
    Then, every student receives (weakly) higher expected utility under paired serial dictatorship than under independent RSD. 
\end{proposition}

The proof of Proposition \ref{ex_ante_improvement_1} relies on student signals depending \textit{only} on relative preferences. As a result, students can ``mimic'' their independent RSD allocation. In the general case, different students may choose to reveal a higher signal to compete for different courses, making analysis less tractable. If we additionally ignore the combinatorial nature of course allocation (e.g. jointly allocating study abroad slots and outdoor education trips or over-competitive ``Social Dance'' classes and dorms), we can get an even stronger result:

\begin{theorem}[Ex-Ante Efficiency]\label{ex_ante_efficienct_1}
    Suppose that in addition to common values, students only demand one course and one dorm so 
    $$u_i(c, d) = \frac{\lambda_i}{1+\lambda_i}v(c) + \frac{1}{1+\lambda_i}w(d).$$
    Then, for any signal space $\mcal S$, paired serial dictatorship is ex-ante optimal among all mechanisms that have signal space $\mcal S$.
\end{theorem}

Unfortunately, the proof of Theorem \ref{ex_ante_efficienct_1} relies strongly on course allocation not being combinatorial. We set up the problem as an optimal transport problem between (1) student relative preferences and signals, (2) signals and courses, and (3) signals and dorms. With the appropriate orders, each of the three problems has a supermodular objective and one-dimensional spaces, which implies that the optimal transport plan is co-monotone. In the combinatorial case, the set of subsets of $\mcal C$ is no longer one-dimensional, and the join of two schedules of desirable courses may force a student into taking too many courses and thus be hazardous to their utility. Similarly, gains in utility can be found by taking an insignificant course from a student who chooses courses early and giving it to a student who chooses courses later and values it more. As such, optimality of the co-monotone transport is lost.

\section{Simulations}\label{simulations}

To quantify the realized differences between paired serial dictatorship and the current mechanism, we simulate both mechanisms in an economy with $1,000$ students, each choosing four of $40$ courses and one of $10$ dorms. Let $\mcal C = \{0, ..., 39\}$ and $\mcal D = \{0, ..., 9\}$. Student preferences were parameterized so student $i$ has
$$u_i(C, d) = \lambda_i \sum_{c \in C} (v_c^i + 0.025c) + \gamma_i (w_d^i + 0.1d)$$
where $\lambda_i, \gamma_i \sim U(0, 10), v_c^i, w_d^i \sim U(0, 5)$ and $U(a, b)$ denotes the uniform distribution over the interval $[a, b]$. The $0.025c$ and $0.1d$ represent some common value for certain courses and dorms (all students prefer courses or a dorm with a higher label), while $v_c^i, w_d^i$ represent idiosyncratic preferences. Students are restricted to choose up to four courses and one dorm. Course and dorm quantities are initialized to be a random partition of the total quantity of courses or dorms demanded. The signal space is $\mcal S = \{0, 1, ..., 9\}$ with the usual ordering. 

As the selection of courses and dorms after orders are determined is strategyproof, we only need to learn an equilibrium of the induced game where each student's strategy space is $\mcal S$. To learn equilibrium, we use exponential weight updating (EXP3), a no-regret learning algorithm which converges to a coarse correlated equilibrium (see \cite{fudenberg_2009_learning} or \cite{hart_2013_simple} for details). Initially, each student has a uniform random distribution over signals. In each iteration, each student draws a signal, observes their payoff from that run of the game and their counterfactual utility had they played each other signal, and increases the probability of playing an action the student regrets not playing in the current iteration for the next iteration. The algorithm runs for 200 iterations, with each iteration utilizing 200 draws of $r^c, r^d$. Learning dynamics are plotted in Figure \ref{regret_10signals}: 

\begin{figure}[h]
    \centering
    \includegraphics[width=0.9\linewidth]{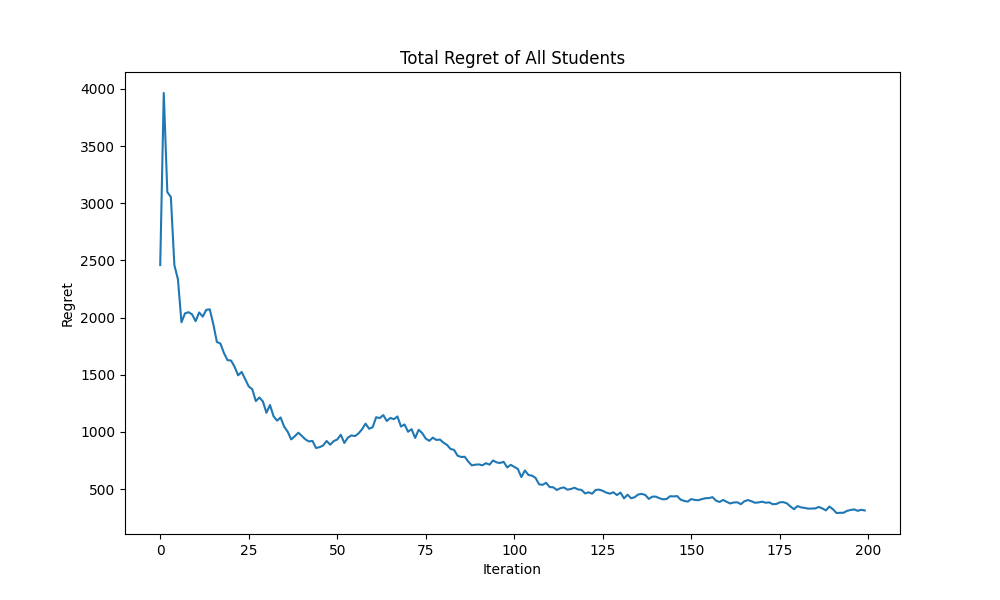}
    \caption{Regret over Learning Iterations}
    \label{regret_10signals}
\end{figure}

While EXP3 only guarantees convergence to a coarse correlated equilibrium, $993$ (out of $1000$) students play their most likely signal with probability greater than $0.9$, while $967$ students play their most likely signal with probability at least $1-10^{-9}$. As a correlated equilibrium where students only play a single strategy is a Nash equilibrium (in pure strategies), the learned equilibrium is a good approximation of Nash equilibrium as well. While there may be multiple Nash equilibria of the game induced by paired serial dictatorship, EXP3 often finds a coarse correlated equilibrium which mixes over the multiplicity of Nash equilibria. Such mixing not being observed here suggests that either (1) there is a unique equilibrium or (2) the Nash equilibria are similar for most agents and equilibrium selection is not a key driver of welfare. Additionally, the algorithm converges quite quickly: From iteration 175 onward, total regret holds steady at a low level. In many other settings, EXP3 requires over an order of magnitude more iterations. This hints to paired serial dictatorship being a ``simple'' mechanism.

Next, we plot students' equilibrium signal against the logarithm of their ratio of course preference to dorm preference, $\log(\lambda_i / \gamma_i)$ as the logarithm of ratios is a symmetric distribution but the distribution of ratios itself has a large right tail. Results are in Figure \ref{signals_10signals}:

\begin{figure}[h]
    \centering
    \includegraphics[width=1\linewidth]{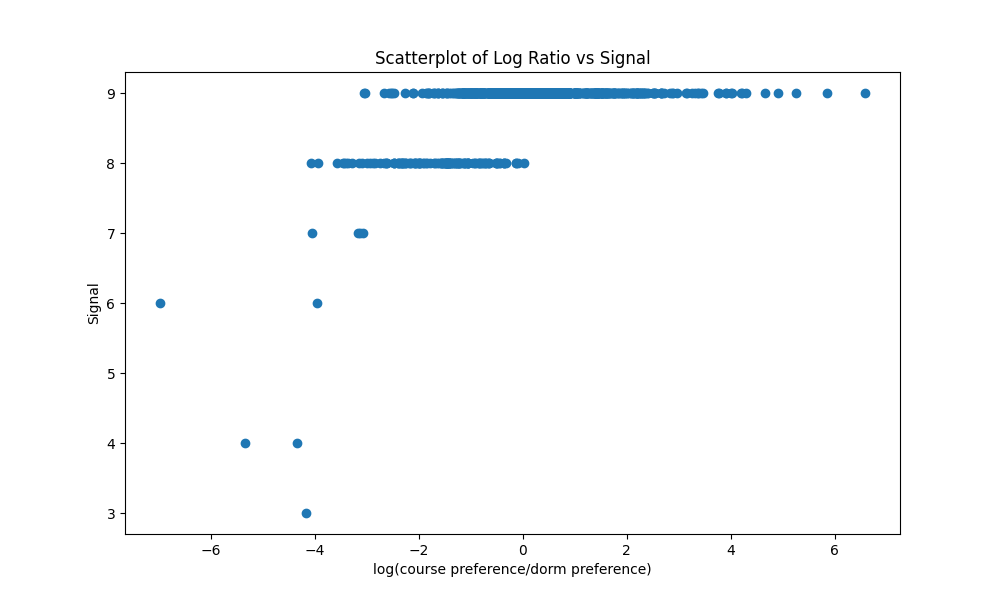}
    \caption{Student Equilibrium Signals}
    \label{signals_10signals}
\end{figure}

In this particular setting, students exhibit bunching at high signals, suggesting that good courses are more competitive than good dorms. One hypothesis as to why this is the case is that course selection determines which \textit{four} courses are chosen while dorm selection only determines a student's one dorm. This force can be calibrated by changing the common value component of students' utilities from courses or dorms. Students with higher log ratios tend to report higher signals, corroborating the intuition behind Proposition \ref{preferences_seperation}. However, Proposition \ref{preferences_seperation} does not perfectly apply, as students generally will not care more about the same courses or dorms than another student due to pseudo-random initialization of preferences.

Next, we consider changes in student welfare. As student utilities were initialized independent of one another, it is possible for one student to have twice as much 'utils' from any bundle as another student, so comparisons of aggregate utility or even changes in utility are not meaningful in this setting. However, we can derive meaningful comparisons from analyzing each student's percent change in welfare between the two mechanisms. Each individual student's expected utility and standard deviation in utilities (across $r^c, r^d$ draws) were generally improved. Out of $1,000$ students, $917$ saw an increase in expected utility while $985$ saw a decrease in standard deviation in utility. On average, student expected utility increased by $3.18\%$ while student standard deviations in utilities decreased by $57.28\%$. Histograms of percent changes in a student's expected utility and standard deviations in utility from paired serial dictatorship versus independent RSD are plotted in Figure \ref{joint_indep}:

\begin{figure}[h]
    \centering
    \includegraphics[width=1\linewidth]{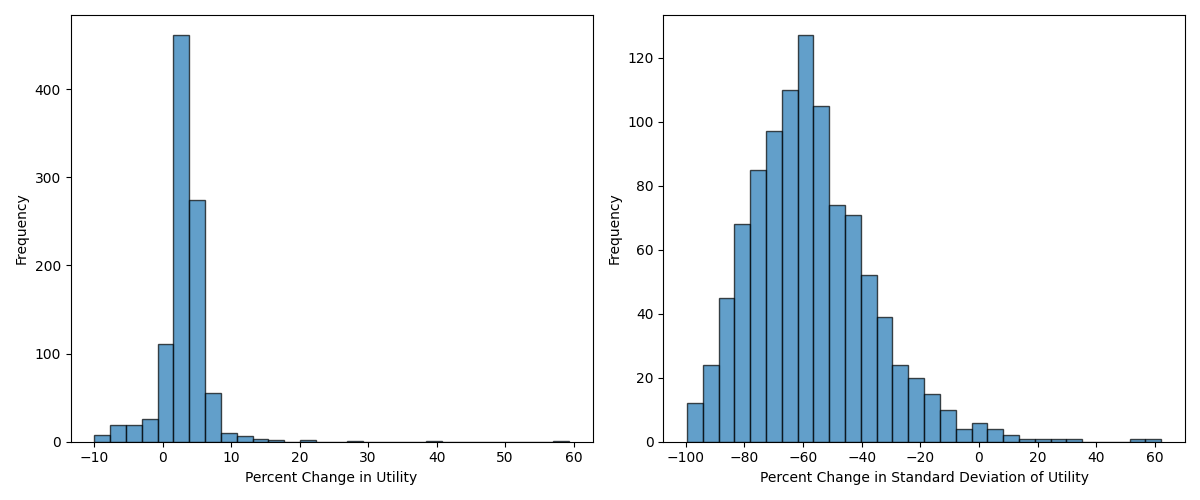}
    \caption{Percent change in measures when moving to paired serial dictatorship from independent RSD}
    \label{joint_indep}
\end{figure}

\subsection{Expanding the Signal Space}

What happens when students have access to a richer set of signals? In general, this can only help students better express preferences. To test this, we keep the same randomly initialized environment as before and only change the signal space to be $\{0, 1,..., 19\}$. Learning dynamics in this setting are similar to before: Total regret over time is comparable and a majority of students converge to basically a pure strategy: $970$ students play their most likely signal with probability greater than $0.9$, while $852$ students play their most likely signal with probability at least $1-10^{-9}$. Since students simply have twice as many signals as before, learning is slower in this setting. Equilibrium behavior still exhibits more competition for courses than dorms, but more separation is observed with students pooling between three signals (17, 18, 19) instead of the previous two (8, 9). Compared to before, more students also play lower signals outside of the signals that other students pool on. Note that Proposition \ref{signal_expansion} does not apply since the original equilibrium did not induce a deterministic allocation and exponential weight updating may not learn the same equilibria if there are multiple. Random sampling of $r^c, r^d$ between trials and iterations may also lead to different equilibria being learned. Student behavior is plotted in Figure \ref{signals_20signals}:

\begin{figure}[h]
    \centering
    \includegraphics[width=1\linewidth]{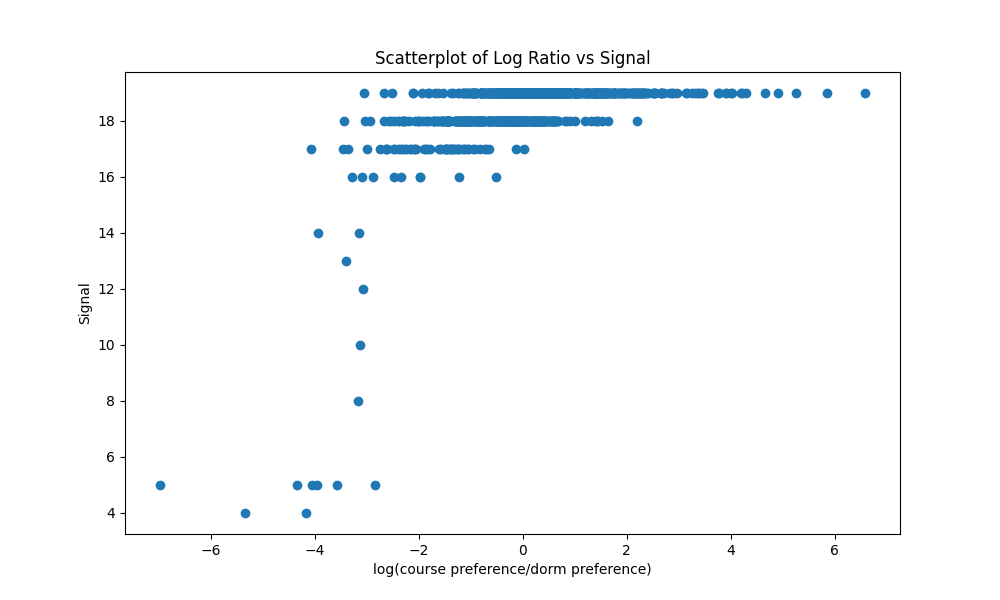}
    \caption{Student Equilibrium Signals (Expanded Signal Space)}
    \label{signals_20signals}
\end{figure}

Even though we are not at a deterministic allocation, the fewer students that pool on any one given signal, the less randomization plays a role. Not only does this imply a better resolution of coordination frictions, but there is also less variation in each student's allocation when tie-breaking plays less of a role. This intuition is similar to a continuous version of Theorem \ref{deterministic_pareto}. The gains from increasing the signal space from 10 to 20 are much more modest than the gains from adopting paired serial dictatorship over independent RSD. Out of $1000$ students, expected utility increased for $691$ students and standard deviations in utility decreased for $807$ students. On average, expected utility increased by $0.86\%$ and standard deviations in utility decreased by $16.36\%$. Histograms of percent changes in a student’s expected utility and standard deviations in utility from paired serial dictatorship with $20$ signals versus paired serial dictatorship with $10$ signals are in Figure \ref{20s_10s}:

\begin{figure}[h]
    \centering
    \includegraphics[width=1\linewidth]{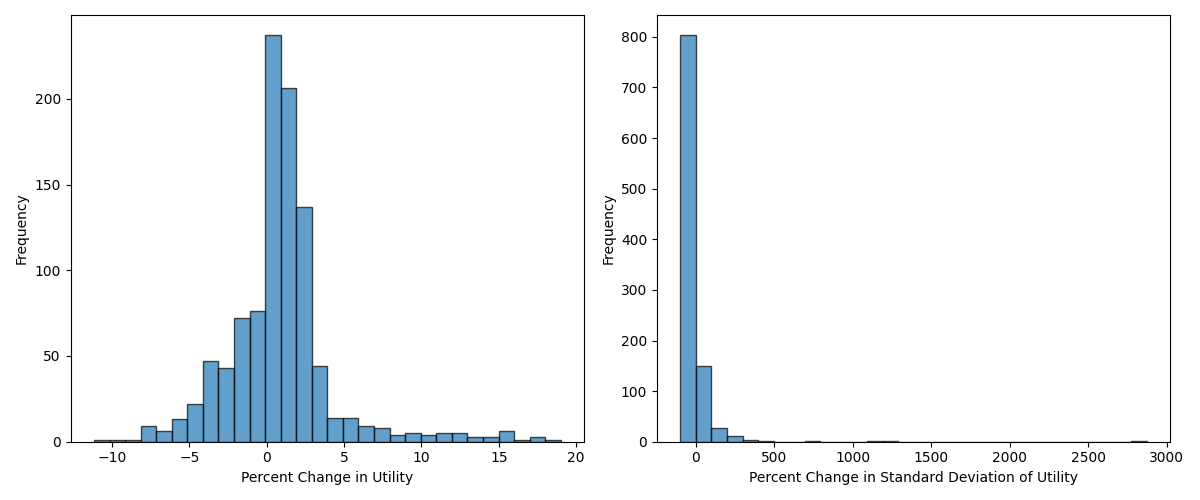}
    \caption{Percent change in measures under paired serial dictatorship when moving to 20 signals from 10 signals.}
    \label{20s_10s}
\end{figure}

\subsection{Tie-breaking Before Signaling}

Another possible adaptation to paired serial dictatorship would be to draw $r^c, r^d$ first and let students observe their tie-breaking number before deciding what signal to reveal. To test this (back in the case when there are only ten signals), we first randomly sample $200$ draws of $r^c, r^d$ and learn equilibria induced by those tie-breaking orders. This is once again done with the same randomly initialized environment as before. Unfortunately, since signal now depends on tie-breaking, a plot of student equilibrium signals is no longer well defined; looking into measures to see if this change decreased pooling would be of interest. However, this change is generally \textit{not} beneficial: $868$ students see a reduction in expected utility and $948$ students see an increase in standard deviation in utility. On average, expected utility decreases by $1.23\%$ and standard deviation in utility increases by $77.32\%$ compared to tie-breaking after signals are reported. Histograms of percent changes in a student’s expected utility and standard deviations in utility from paired serial dictatorship with tie-breaking before signal selection versus after signal selection are plotted in Figure \ref{before_after}:
\begin{figure}[H]
    \centering
    \includegraphics[width=1\linewidth]{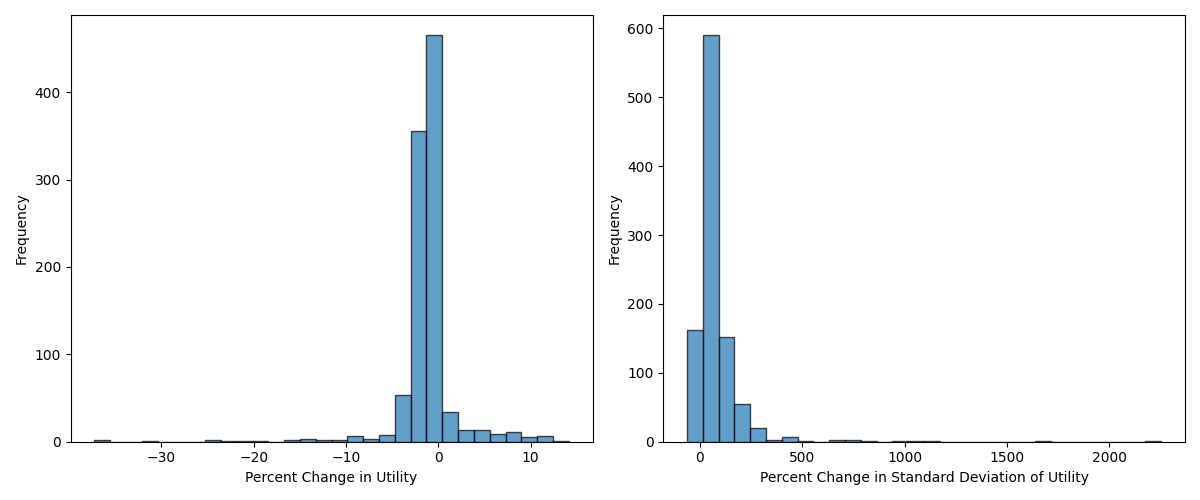}
    \caption{Percent change in measures under paired serial dictatorship when moving to tie-breaking before students select signals from tie-breaking after students select signals.}
    \label{before_after}
\end{figure}

Even though students' strategies can condition on $r^c$ and $r^d$ leading to the allocation being deterministic, the way $r^c$ and $r^d$ are assigned is still uncorrelated with students' relative preferences. As such, the key force leading deterministic allocations being efficient is not the fact that they are deterministic in and of themselves, but rather that they fully resolve coordination frictions. However, running paired serial dictatorship where ties are broken before students signal relative preferences is still an improvement over independent RSD, as students still have an opportunity to signal relative preferences.

\subsection{Bounded Rationality}

One final consideration is that students may not perfectly play equilibrium actions. To compute what signal to reveal in equilibrium, each student would need to know not only the distribution of other students' relative preferences, but also their utility from bundles of courses and dorms. In any moderately sized problem, this computation is also generally intractable. Instead, consider the case in which students were myopic and simply reported a signal based on their own relative preferences, regardless of greater strategic concerns. To test this, we suppose student $i$ plays
$$\hat{s}(i) = \argmin_{s \in \{0, 1, ..., 9\}} |4.5+\log(\lambda_i / \gamma_i) - s|$$
where $4.5$ is halfway between the smallest signal $0$ and the largest signal $9$ and $\log(\lambda_i / \gamma_i)$ is symmetrically distributed around $0$ with support approximately $[-6.5, 6.5]$. Alternatively, we can think of this case as a heuristic test of Proposition \ref{ex_ante_improvement_1} when the common value assumption fails. In the simulation setup, students do not have common values for courses and dorms, but under this specification of bounded rationality, students behave as if the common value assumption holds: Relative preferences are the only force driving students' signals (formalized in Proposition \ref{preferences_seperation}) while the competitiveness of different courses or dorms (formalized in Proposition \ref{competition_seperation}) is ignored.

Even under bounded rationality, paired serial dictatorship outperforms independent RSD: 869 students see an increase in expected utility while all 1000 students have lower standard deviation in utilities. On average, student expected utilities increase by $4.54\%$ while standard deviations in utilities decreased by $84.36\%$. In particular, 35 students receive a deterministic bundle. These measures are depicted in Figure \ref{bounded_rationality}:

\begin{figure}[H]
    \centering
    \includegraphics[width=1\linewidth]{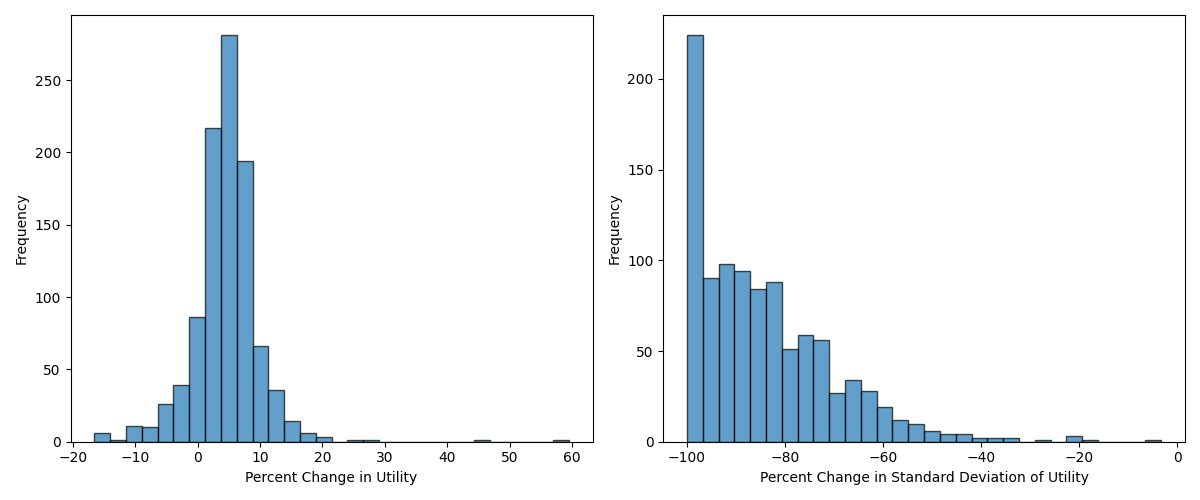}
    \caption{Percent change in measures when moving to paired serial dictatorship with bounded rationality from independent RSD.}
    \label{bounded_rationality}
\end{figure}

\section{Conclusion}\label{conclusion}
This paper introduced and analyzed paired serial dictatorship, a new mechanism for the allocation of multiple goods in the absence of transfers. By linking previously disjoint one-sided matching markets, priority in one market can effectively be used as a shadow price on the other market to screen agents' relative preferences between the goods offered in each market. Although serial dictatorship is efficient within each market, paired serial dictatorship can still achieve improvements in the aggregate. In cases where the equilibrium allocation is deterministic, paired serial dictatorship guarantees ex-post Pareto efficiency, a significant improvement over the inefficiency random tie-breaking induces in independent RSD. In other cases where students differ only in their relative preferences between courses and dorms, paired serial dictatorship is an ex-ante improvement over independent RSD. Simulations suggest that paired serial dictatorship leads to higher expected utility and less uncertainty even when the prior two assumptions do not hold. The mechanism also remains effective under bounded rationality, suggesting robustness in practical implementation. It is furthermore a minimal change over current independent RSD mechanisms that are widespread at many universities. 

Several open questions remain for future research. First, identifying conditions on student preferences or the underlying environment that guarantee the existence of a deterministic equilibrium allocation can clarify when randomization is necessary in good allocation problems. In particular, suppose that instead of dorms, one side of the market consisted of bundles of cash. If the designer can freely vary the amount of numeraire in each bundle to induce a deterministic equilibrium allocation, then those values of numeraire can be thought of as market-clearing prices in a combinatorial auction setting. Similarly, finding weaker conditions that fully characterize when paired serial dictatorship is an ex-ante improvement over independent RSD is also of interest. 

Second, additional simulations with different specifications can provide a richer understanding of the performance of paired serial dictatorship. For example, the environment considered in this paper was fairly balanced between courses and dorms while most students had similar relative preferences between courses and dorms. Perhaps in real-world course and dorm allocation settings, the distribution of relative preferences is bimodal or non-symmetric. Unfortunately, since courses and dorms have generally been allocated separately, there is no way to recover relative preferences using currently existing data. An experimental approach or pilot tests can be used to address this gap; this may have broader insights into implementation and how agents strategically interact with the mechanism (or fail to).

Third, formalizing notions of how (and if) paired serial dictatorship is ``simple'' to learn or evaluating how welfare changes when students have limited rationality would inform the implementation of this mechanism. Another simulation to run would be to explore dynamics when some subset of the population was myopic while the remainder learned an equilibrium (conditional on myopic players having some fixed action). In a similar vein, strategyproofness is only defined in cases where each agent's signal space is equal to their type space. In some allocation problems, it may be infeasible for agents to fully report their type (i.e. it is computationally infeasible to fully report combinatorial preferences). If strategyproofness is a primary concern for the designer, paired serial dictatorship can always be implemented by the direct incentive compatible mechanism induced via the revelation principle. Under this implementation, other students' reports effectively generate a menu for $i$, leading to truthful reporting being a dominant strategy. Developing notions of strategyproofness more broadly in such settings is of interest. 

Finally, paired serial dictatorship can be extended in a variety of directions. Most notably is generalizing to an arbitrary number of markets. In this case, we can interpret one-dimensional signals more generally as how much of a fictitious currency students ``spend'' on a higher priority in each market, subject to a budget constraint. When allocating courses and dorms, students need to choose courses for each of the multiple quarters or semesters in the year; comparing one signal determining course priority in every term to interpreting each term as a different market and running this extension of paired serial dictatorship would be of interest. Another practical concern is that many students choose dorms in groups to ensure they can be roommates. Investigating how paired serial dictatorship can be adapted to include such preferences is another potential direction. Third, random serial dictatorship is not the only mechanism that requires randomization: What if other mechanisms that involve randomization were paired? In particular, the approximate CEEI mechanism that is used in practice at some schools requires randomization in students' initial budgets to ensure a market-clearing allocation exists. Allowing students to report signals that determine their dorm selection priority and their initial budget under CEEI is a possibility. 

Beyond the application of paired course and dorm allocation discussed in this paper, paired serial dictatorship may be applicable in a variety of other settings. For example, when allocating different forms of support for low-income communities, paired serial dictatorship may be used to elicit relative preferences between public housing units and priority for food shelter donations. Similarly, joining public school and daycare/after-school program admissions can also lead to improvements in welfare. Paired serial dictatorship can also be used within organizations to allocate limited resources like parking spots and meeting rooms. Implementation in any of these settings can resolve inefficiencies generated by ad-hoc tie-breaking being uncorrelated with relative preferences.

\newpage

\bibliography{cites}

@article{budish_2010_finding,
  author = {Budish, Eric and Sandholm, Tuomas and Othman, Abraham},
  month = {05},
  pages = {873-880},
  title = {Finding approximate competitive equilibria: efficient and fair course allocation},
  doi = {https://dl.acm.org/doi/10.5555/1838206.1838323},
  urldate = {2024-06-26},
  year = {2010},
  journal = {Adaptive Agents and Multi-Agents Systems}
}

@article{budish_2011_the,
  author = {Budish, Eric},
  month = {12},
  pages = {1061-1103},
  title = {The Combinatorial Assignment Problem: Approximate Competitive Equilibrium from Equal Incomes},
  doi = {10.1086/664613},
  urldate = {2020-10-03},
  volume = {119},
  year = {2011},
  journal = {Journal of Political Economy}
}

@article{budish_2012_the,
  author = {Budish, Eric and Cantillon, Estelle},
  month = {08},
  pages = {2237-2271},
  title = {The Multi-unit Assignment Problem: Theory and Evidence from Course Allocation at Harvard},
  doi = {10.1257/aer.102.5.2237},
  urldate = {2020-04-02},
  volume = {102},
  year = {2012},
  journal = {American Economic Review}
}

@article{schmeidler_1973_equilibrium,
  author = {Schmeidler, David},
  month = {04},
  pages = {295-300},
  publisher = {Springer Science+Business Media},
  title = {Equilibrium Points of Nonatomic Games},
  doi = {10.1007/bf01014905},
  urldate = {2024-09-08},
  volume = {7},
  year = {1973},
  journal = {Journal of Statistical Physics}
}

@article{milgrom_1994_monotone,
  author = {Milgrom, Paul and Shannon, Chris},
  month = {01},
  pages = {157},
  title = {Monotone Comparative Statics},
  doi = {10.2307/2951479},
  volume = {62},
  year = {1994},
  journal = {Econometrica}
}

@article{horner_2015_dynamic,
  author = {Horner, Johannes and Guo, Yingni},
  title = {Dynamic Mechanisms Without Money},
  doi = {10.2139/ssrn.2563005},
  urldate = {2019-10-08},
  year = {2015},
  journal = {SSRN Electronic Journal}
}

@article{balseiro_2019_multiagent,
  author = {Balseiro, Santiago R. and Gurkan, Huseyin and Sun, Peng},
  month = {09},
  pages = {1417-1436},
  title = {Multiagent Mechanism Design Without Money},
  doi = {10.1287/opre.2018.1820},
  urldate = {2023-01-01},
  volume = {67},
  year = {2019},
  journal = {Operations Research}
}

@article{carmona_2020_pure,
  author = {Carmona, Guilherme and Podczeck, Konrad},
  month = {05},
  pages = {105015},
  title = {Pure strategy Nash equilibria of large finite-player games and their relationship to non-atomic games},
  doi = {10.1016/j.jet.2020.105015},
  urldate = {2021-05-17},
  volume = {187},
  year = {2020},
  journal = {Journal of Economic Theory}
}

@article{bade_2020_random,
  author = {Bade, Sophie},
  month = {02},
  pages = {353-368},
  title = {Random Serial Dictatorship: The One and Only},
  doi = {10.1287/moor.2019.0987},
  volume = {45},
  year = {2020},
  journal = {Mathematics of Operations Research}
}

@misc{bhalgat_2011_social,
  author = {Bhalgat, Anand and Chakrabarty, Deeparnab and Khanna, Sanjeev},
  title = {Social Welfare in One-Sided Matching Markets without Money},
  url = {https://www.cis.upenn.edu/~sanjeev/papers/approx11_matching.pdf},
  urldate = {2024-10-29},
  year = {2011}
}

@article{jackson_2007_overcoming,
  author = {Jackson, Matthew O and Sonnenschein, Hugo F},
  month = {01},
  pages = {241-257},
  title = {Overcoming Incentive Constraints by Linking Decisions},
  doi = {10.1111/j.1468-0262.2007.00737.x},
  urldate = {2022-03-08},
  volume = {75},
  year = {2007},
  journal = {Econometrica}
}

@article{noda_2024_no,
  author = {Noda, Shunya and Okada, Genta},
  month = {08},
  publisher = {Cornell University},
  title = {No Screening is More Efficient with Multiple Objects},
  doi = {10.48550/arxiv.2408.10077},
  urldate = {2024-11-02},
  year = {2024},
  journal = {arXiv (Cornell University)}
}

@article{kornbluth_2021_undergraduate,
  author = {Kornbluth, Daniel and Kushnir, Alexey I.},
  title = {Undergraduate Course Allocation through Pseudo-Markets},
  doi = {10.2139/ssrn.3901146},
  urldate = {2022-08-11},
  year = {2021},
  journal = {SSRN Electronic Journal}
}

@misc{kesten_2021_strategyproof,
  author = {Kesten, Onur and Selcuk, Ozyurt},
  title = {Strategy-proof and Efficient Mediation: An Ordinal Market Design Approach},
  url = {https://www.econ.iastate.edu/files/events/files/56c471_3e438518e8634f68819db0d0bb64f129.pdf},
  urldate = {2024-11-02},
  year = {2021}
}

@article{svensson_1999_strategyproof,
  author = {Svensson, Lars-Gunnar},
  month = {11},
  publisher = {RELX Group (Netherlands)},
  title = {Strategy-Proof Allocation of Indivisible Goods},
  urldate = {2024-11-02},
  year = {1999},
  journal = {SSRN Electronic Journal}
}

@article{chakravarty_2013_optimal,
  author = {Chakravarty, Surajeet and Kaplan, Todd R.},
  month = {01},
  pages = {1-20},
  title = {Optimal allocation without transfer payments},
  doi = {10.1016/j.geb.2012.08.006},
  volume = {77},
  year = {2013},
  journal = {Games and Economic Behavior}
}

@article{basteck_2024_an,
  author = {Basteck, Christian},
  month = {01},
  publisher = {RELX Group (Netherlands)},
  title = {An Axiomatization of the Random Priority Rule},
  doi = {10.2139/ssrn.4780639},
  urldate = {2024-11-02},
  year = {2024},
  journal = {SSRN Electronic Journal}
}

@article{imamura_2024_efficient,
  author = {Imamura, Kenzo and Kawase, Yasushi},
  month = {03},
  pages = {197-207},
  publisher = {Elsevier BV},
  title = {Efficient matching under general constraints},
  doi = {10.1016/j.geb.2024.03.013},
  urldate = {2024-11-02},
  volume = {145},
  year = {2024},
  journal = {Games and Economic Behavior}
}

@article{romeromedina_2024_strategic,
  author = {Romero-Medina, Antonio and Matteo Triossi},
  month = {08},
  pages = {106701-106701},
  publisher = {Elsevier BV},
  title = {Strategic priority-based course allocation},
  doi = {10.1016/j.jebo.2024.106701},
  urldate = {2024-11-02},
  volume = {226},
  year = {2024},
  journal = {Journal of Economic Behavior \& Organization}
}

@article{bichler_2021_randomized,
  author = {Bichler, Martin and Merting, Soeren},
  month = {07},
  pages = {3540-3559},
  title = {Randomized Scheduling Mechanisms: Assigning Course Seats in a Fair and Efficient Way},
  doi = {10.1111/poms.13449},
  urldate = {2021-11-27},
  volume = {30},
  year = {2021},
  journal = {Production and Operations Management}
}

@article{kikuchi_2024_a,
  author = {Kikuchi, Kazuya and Koriyama, Yukio},
  month = {03},
  pages = {789-797},
  publisher = {Springer Science and Business Media LLC},
  title = {A general impossibility theorem on Pareto efficiency and Bayesian incentive compatibility},
  doi = {10.1007/s00355-024-01515-4},
  urldate = {2024-11-02},
  volume = {62},
  year = {2024},
  journal = {Social Choice and Welfare}
}

@article{ezzatelokda_2023_a,
  author = {Ezzat Elokda and Saverio Bolognani and Censi, Andrea and Florian Dörfler and Frazzoli, Emilio},
  month = {04},
  pages = {578-610},
  publisher = {Birkhäuser},
  title = {A Self-Contained Karma Economy for the Dynamic Allocation of Common Resources},
  doi = {10.1007/s13235-023-00503-0},
  urldate = {2024-07-04},
  volume = {14},
  year = {2023},
  journal = {Dynamic games and applications}
}

@article{budish_2017_course,
  author = {Budish, Eric and Cachon, Gérard P. and Kessler, Judd B. and Othman, Abraham},
  month = {04},
  pages = {314-336},
  title = {Course Match: A Large-Scale Implementation of Approximate Competitive Equilibrium from Equal Incomes for Combinatorial Allocation},
  doi = {10.1287/opre.2016.1544},
  url = {https://users.nber.org/~kesslerj/papers/Budish_etal_ACEEI_2017.pdf},
  urldate = {2021-05-07},
  volume = {65},
  year = {2017},
  journal = {Operations Research}
}

@book{filipposantambrogio_2015_optimal,
  author = {Filippo Santambrogio},
  month = {10},
  publisher = {Birkhäuser},
  title = {Optimal Transport for Applied Mathematicians},
  year = {2015}
}

@article{luo_2024_marginal,
  author = {Luo, Daniel and Wolitzky, Alexander},
  month = {11},
  title = {Marginal Reputation},
  year = {2024},
  journal = {Preprint}
}

@article{fudenberg_2009_learning,
  author = {Fudenberg, Drew and Levine, David K.},
  month = {09},
  pages = {385-420},
  title = {Learning and Equilibrium},
  doi = {10.1146/annurev.economics.050708.142930},
  volume = {1},
  year = {2009},
  journal = {Annual Review of Economics}
}

@book{hart_2013_simple,
  author = {Hart, Sergiu and Andreu Mas-Colell},
  month = {01},
  publisher = {World Scientific Series in Economic Theory: Volume 4},
  title = {Simple Adaptive Strategies: From Regret-Matching to Uncoupled Dynamics},
  year = {2013}
}

@misc{ball_2023_quota,
  author = {Ball, Ian and Kattwinkel, Deniz},
  title = {Quota Mechanisms: Finite-Sample Optimality and Robustness},
  url = {https://arxiv.org/abs/2309.07363},
  urldate = {2024-12-28},
  year = {2023},
  organization = {arXiv.org}
}

@article{prendergast_2022_the,
  author = {Prendergast, Canice},
  month = {04},
  title = {The Allocation of Food to Food Banks},
  doi = {10.1086/720332},
  year = {2022},
  journal = {Journal of Political Economy}
}

@article{konishi_2001_on,
  author = {Konishi, Hideo and Quint, Thomas and Wako, Jun},
  month = {02},
  pages = {1-15},
  title = {On the Shapley–Scarf economy: the case of multiple types of indivisible goods},
  doi = {10.1016/s0304-4068(00)00061-6},
  urldate = {2021-12-13},
  volume = {35},
  year = {2001},
  journal = {Journal of Mathematical Economics}
}

@article{ppai_2001_strategyproof,
  author = {Pápai, Szilvia},
  month = {07},
  pages = {257-271},
  publisher = {Wiley},
  title = {Strategyproof and Nonbossy Multiple Assignments},
  doi = {10.1111/1097-3923.00066},
  volume = {3},
  year = {2001},
  journal = {Journal of Public Economic Theory}
}

@article{klaus_2007_the,
  author = {Klaus, Bettina},
  month = {11},
  pages = {919-924},
  publisher = {Elsevier},
  title = {The coordinate-wise core for multiple-type housing markets is second-best incentive compatible},
  doi = {10.1016/j.jmateco.2007.05.013},
  url = {https://www.sciencedirect.com/science/article/pii/S0304406807000936},
  volume = {44},
  year = {2007},
  journal = {Journal of Mathematical Economics}
}

@article{feigenbaum_2020_dynamic,
  author = {Feigenbaum, Itai and Yash Kanoria and Lo, Irene and Sethuraman, Jay},
  month = {11},
  pages = {5341-5361},
  publisher = {Institute for Operations Research and the Management Sciences},
  title = {Dynamic Matching in School Choice: Efficient Seat Reassignment After Late Cancellations},
  doi = {10.1287/mnsc.2019.3469},
  volume = {66},
  year = {2020},
  journal = {Management Science}
}

@article{anno_2016_on,
  author = {Anno, Hidekazu and Morimitsu Kurino},
  month = {10},
  pages = {166-185},
  publisher = {Elsevier BV},
  title = {On the operation of multiple matching markets},
  doi = {10.1016/j.geb.2016.10.001},
  urldate = {2025-03-30},
  volume = {100},
  year = {2016},
  journal = {Games and Economic Behavior}
}

@article{nguyen_2024_nearsubstitute,
  author = {Nguyen, Thành and Vohra, Rakesh},
  month = {12},
  pages = {4122-4154},
  title = {(Near-)Substitute Preferences and Equilibria with Indivisibilities},
  doi = {10.1086/731413},
  urldate = {2025-03-29},
  volume = {132},
  year = {2024},
  journal = {Journal of Political Economy}
}

@article{takahashi2010community,
  title={Community enforcement when players observe partners' past play},
  author={Takahashi, Satoru},
  journal={Journal of Economic Theory},
  volume={145},
  number={1},
  pages={42--62},
  year={2010},
  publisher={Elsevier}
}

\newpage

\appendix

\section*{Appendix A: Mechanism Details and Equilibrium}
\makeatletter\def\@currentlabel{Appendix A}\makeatother
\label{Appendix A}

This Appendix provides details about the process of students choosing courses and dorms in step (5), verifies that the process is well-defined, and shows that a pure strategy Nash equilibrium exists in the induced game. First, notation used throughout the appendices:
\begin{itemize}
    \item $e_i(c), e_i(d)$ denotes the mass of students enrolled in a course $c$ or dorm $d$ by the time it is student $i$'s turn to choose. 
    \item $\psi^c, \psi^d$ are (measurable, bijective) re-labelings of students given signals, $r^c, r^d$ so student $i$ chooses courses before student $j$ if and only if $\psi^c(i) > \psi^c(j)$ and dorms before if and only if $\psi^d(i) > \psi^d(j)$.
    \item $\hat{C}(i, s|\hat{s}, r^c)$, $\hat{D}(i, s|\hat{s}, r^d)$ denote the courses and dorms student $i$ gets when sending signal $s$ given others' signals being $\hat{s}$ and tie-breaking rules $r^c, r^d$.
\end{itemize}
Then, a set of courses $C$ (similarly dorms) is available for student $i$ if 
$$e_i(c) < q(c) \text{ for all } c \in C.$$ 
For paired serial dictatorship to be a well-defined process, $e_i(c)$ and $e_i(d)$ need to be well-defined objects, as there may be measurability issues in the set of students who have chosen a certain course at any given point in time.

\begin{proposition}\label{well_def}
    With the measurability assumption (A1), $e_i(c), e_i(d)$ are well defined for any measurable $\hat{s}$.
\end{proposition}

\begin{proof}
    By symmetry, it is sufficient to show that $e_i(c)$ is well defined. We will first re-label students by the order they choose courses, and then show the result by induction on the number of courses that have reached capacity. 

    \noindent \textbf{Step One: Relabeling Students}

    We will work with the particular re-labeling
    $$\psi^c(i) = \frac{\hat{s}(i) + r^c(i) -1}{\max (\mcal S)}.$$
    This re-labeling has the additional attractive property that changes in $\hat{s}$ do not change any particular student $i$'s relabeling $\psi^c(i)$ (unless $\hat{s}(i)$ itself changes); changing $\hat{s}$ only changes the \textit{distribution} of students over $[0,1]$ after $\psi^c$ is applied.
    
    First, $\psi^c$ is measurable (when we also equip the image of $\psi^c$ with the Borel $\sigma$-algebra). We will show that $(\psi^c)^{-1}(a,b)$ is measurable for any $a, b \in [0,1]$. We can write
    \begin{align*}
        & (\psi^c)^{-1}((a,b)) \\
        =& \{i: i \text{ chooses between } (\psi^c)^{-1}(a) \text{ and } (\psi^c)^{-1}(b)\} \\
        =& \{i: \hat{s}((\psi^c)^{-1}(a)) < \hat{s}(i) < \hat{s}((\psi^c)^{-1}(b))\} \\
        & \cup \{i: \hat{s}((\psi^c)^{-1}(a)) = \hat{s}(i), r^c((\psi^c)^{-1}(a)) < r^c(i)\} \\
        & \cup \{i: \hat{s}((\psi^c)^{-1}(b)) = \hat{s}(i), r^c(i) < r^c((\psi^c)^{-1}(b))\} \\
        =& \hat{s}^{-1}(\{s \in \mcal S: \hat{s}((\psi^c)^{-1}(a)) < s < \hat{s}((\psi^c)^{-1}(b))\}) \\
        & \cup \left[\hat{s}^{-1}(\hat{s}((\psi^c)^{-1}(a)) \cap \{i: r^c((\psi^c)^{-1}(a)) < r^c(i) \right] \\
        & \cup \left[\hat{s}^{-1}(\hat{s}((\psi^c)^{-1}(b)) \cap \{i: r^c(i) < r^c((\psi^c)^{-1}(b)) \right].
    \end{align*}
    By assumption, each of the sets above are measurable. As the finite intersection and union of measurable sets is measurable, we have that $(\psi^c)^{-1}((a,b))$ is measurable. 
    
    As such, we can now work with the pushforward $\nu = \mu \circ (\psi^c)^{-1}$ on the image of $\psi^c$ in Step Two, where students are now ordered properly.

    \noindent \textbf{Step Two: Inducting on Full Courses}

    In the base case, suppose no course has reached capacity. Let $C_i(\mcal C) = \argmax_{C \in 2^{\mcal C}} v_i(C)$ be the set of courses student $i$ chooses when the set of available courses is $\mcal C$. Let 
    $$i_{c_1} = \sup \left\{i: \min_{c \in \mcal C} \Big(q(c) - \nu(\{j < i: c \in C_j(\mcal C)\}) \Big) > 0\right\}$$
    denote the latest student for which no course has reached capacity yet. This is well-defined as $\mcal C$ is a finite set and for each $c, \{j < i: c \in C_j(\mcal C)\} = [0,i) \cap \{j: c \in C_j(\mcal C)\}$, and $[0,1)$ is measurable while 
    \begin{align*}
        \{j: c \in C_j(\mcal C)\} &= \bigcup_{C \subset \mcal C: c \in C} \{j: v_j(C) \geq v_j(C') \text{ for all } C' \in \mcal C\} \\
        &= \bigcup_{C \subset \mcal C: c \in C} \left(\bigcap_{C' \subset \mcal C} \{j: v_j(C) \geq v_j(C')\}\right)
    \end{align*}
    is a finite union of finite intersections of measurable sets since each $j: v_j(C) \geq v_j(C')$ is measurable by the regularity condition on preferences. Note that when $i_{c_1}$ chooses, there must exist some course $c_1$ such that $\nu(\{j < i_{c_1}: c \in C_j(\mcal C)\}) = q(c)$.\footnote{If there are multiple such courses, any selection works.} By construction, $\nu(\{j < i_{c_1}: c \in C_j(\mcal C)\}) \leq q(c)$ for all $c$. If there were a strict inequality, let 
    $$\epsilon = \min_{c \in \mcal C} \Big(q(c) - \nu(\{j < i_{c_1}: c \in C_j(\mcal C)\}) \Big) > 0.$$
    Then, by the time student $i_{c_1} + \epsilon/2$ chose, 
    $$\min_{c \in \mcal C} \Big(q(c) - \nu(\{j < i_{c_1} + \epsilon/2: c \in C_j(\mcal C)\}) \Big) \geq \epsilon - \epsilon/2 > 0$$
    which is once again a contradiction with how $i_{c_1}$ was constructed. As such, $i_{c_1}$ can also be interpreted as the first student for whom $c_1$ is unavailable.
    
    For any $i \leq i_{c_1}$,\footnote{Strictly speaking, this construction only allows us to make this claim for $i < i_{c_1}$, but $\{i_1\}$ is of measure zero so adding them does not change $e_i$ nor lead to measurability issues.} we then have that
    $$e_i(c) = \nu(\{j < i: c \in C_j(\mcal C)\}).$$
    This concludes the base case.

    Inductively, given $i_{c_1},...,i_{c_n}, c_1,...,c_n$, and $e_i(c)$ for all $c, i \leq i_n$, define $i_{c_{n+1}}$ and $c_{n+1}$ as:
    $$i_{c_{n+1}} = \sup \left\{i: \min_{c \in \mcal C \setminus \{c_1,...,c_n\}} \left(q(c) - e_{i_n}(c) - \nu\Big(\{i_{c_n} \leq j < i: c \in C_j(\mcal C \setminus \{c_1,...,c_n\})\Big) \right) > 0\right\}$$
    and $c_{n+1}$ to be the course in $\mcal C\setminus \{c_1,...,c_n\}$ that has zero excess capacity after student $i_{c_{n+1}}$ chooses. Existence of $i_{c_{n+1}}, c_{n+1}$ follows from similar arguments as before, noting that
    \begin{align*}
        \{j: c \in C_j(\mcal C \setminus \{c_1,...,c_n\})\} &= \bigcup_{C \subset \mcal C \setminus \{c_1,...,c_n\}: c \in C} \{j: w_j(C) \geq w_j(C') \forall C' \subset \mcal C\setminus \{c_1,...,c_n\}\} \\
        &= \bigcup_{C \subset \mcal C \setminus \{c_1,...,c_n\}: c \in C} \left(\bigcap_{C' \subset \mcal C\setminus \{c_1,...,c_n\}} \{j: v_j(C) \geq v_j(C')\}\right)
    \end{align*}
    is still a finite union of a finite intersection of measurable sets, and is hence still measurable. In general, $c_n$ is the $n$th course to reach capacity, and $i_{c_n}$ is the first student for which $c_n$ (and by extension, $c_1,...,c_{n-1}$) is unavailable.

    To finish, we have that for any $i \in [i_{c_n}, i_{c_{n+1}}]$ and $c \in \mcal C$,
    $$e_i(c) = e_{i_{c_n}}(c) + \nu\Big(\{i_n \leq j < i: c \in C_j(\mcal C \setminus \{c_1,...,c_n)\}\Big).$$
    This completes the inductive step, and hence the proof.
\end{proof}

\begin{proposition}
    There exists a pure strategy Nash equilibrium of the induced game.
\end{proposition}

\begin{proof}
    The proof proceeds in four steps: Adapting our model to the framework of \cite{schmeidler_1973_equilibrium} and then verifying the three assumptions of \cite{schmeidler_1973_equilibrium}, which guarantees the existence of a pure strategy Nash equilibrium. 

    \noindent \textbf{Step One: Setup}
    
    The set of players is still $\mcal I = [0,1]$. To anonymize the game, each player chooses from the set of actions
    $$A = \mcal S \times \{f: 2^{\mcal C} \to 2^{\mcal C} \text{ s.t. } f(C) \subseteq C\} \times \{g: 2^{\mcal D} \to 2^{\mcal D} \text{ s.t. } g(D) \subseteq D\}$$
    where $\mcal S$ still represents the set of signals. A function $f$ represents a complete, contingent plan of selecting courses from any set of available courses and similarly, $g$ represents a complete, contingent plan of selecting dorms from any set of available dorms. As $\mcal S, \mcal C, \mcal D$ are all finite, the set $A$ is finite as well.

    An $\mcal I$-strategy is a measurable function $\hat{x}: \mcal I \to \Delta (A)$, where $\Delta (A)$ is viewed as a subset of $\mathbb{R}^{|A|}$. Let $\hat{A}$ denote the set of all $\mcal I$-strategies. Given an action $a \in A$, $\mcal I$-strategy $\hat{x}$, and tie-breaking rules $r^c, r^d$, let $\hat{C}(i, a| \hat{x}, r^c)$ be the set of courses player $i$ gets from playing $a$ when facing $\hat{x}, r^c$. Similarly, let $\hat{D}(i, a| \hat{x}, r^d)$ be the set of dorms player $i$ gets from playing $a$ when facing $\hat{x}, r^d$. Define the auxiliary utility function $U: \mcal I \times \hat{A} \to \mathbb{R}^{|A|}$ by
    $$U_a(i, \hat{x}) =  \E{v_i\Big(\hat{C}(i, a| \hat{x}, r^c) \Big) + w_i\Big( \hat{D}(i, a| \hat{x}, r^d) \Big)}$$
    where $U_a(i, \hat{x})$ is the $a$-th component of $U(i, \hat{x})$ and the expectation is taken with respect to $r^c, r^d \sim R$.

    \noindent \textbf{Step Two: Anonymity}

    \begin{claim}
        Let $\hat{x}_a$ denote the $a$-th component of $\hat{x}$. The auxiliary utility function $U$ depends only on $\int_{\mcal I} \hat{x}_a(i) d\mu(i)$ for $a \in A$.
    \end{claim}

    \begin{proof}
        For each $i$, $\hat{x}(i)$ already encodes information about the signal $i$ reports, as well as a complete, contingent plan of how $i$ selects courses. Furthermore, tie-breaking is done uniformly at random and independent of $i$'s actions, so for any $i, j$ we have that $r^c(i), r^c(j)$ are identically distributed (similarly for $r^d(i), r^d(j)$).
    \end{proof}

    \noindent \textbf{Step Three: Measurability}
    
    \begin{claim}
        For all $\hat{x} \in \hat{A}$ and $a, b \in A$, the set $\{i \in \mcal I: U_a(i, \hat{x}) > U_b(i, \hat{x})\}$ is measurable.
    \end{claim}

    \begin{proof}
        As tie-breaking is done independently of any player's actions, Assumption (A2) gives that any two students who both play $a$ or both play $b$ will receive the same distribution over courses and dorm. Thus, it suffices to show that for any $p^c, \rho^c \in \Delta (2^{\mcal C})$ and $p^d, \rho^d \in \Delta (2^{\mcal D})$, 
        $$\left\{i \in \mcal I: \sum_{C \subseteq \mcal C} p^c(C) \cdot v_i(C) +  \sum_{D \subseteq \mcal D} p^d(D) \cdot w_i(D) > \sum_{C \subseteq \mcal C} \rho^c(C) \cdot v_i(C) +  \sum_{D \subseteq \mcal D} \rho^d(D) \cdot w_i(D)\right\}$$
        is measurable. Fixing $p^c, \rho^c, p^d, \rho^d$, define the function $f$ by
        $$f(i|p^c, \rho^c, p^d, \rho^d) = \sum_{C \subseteq \mcal C} p^c(C) \cdot v_i(C) +  \sum_{D \subseteq \mcal D} p^d(D) \cdot w_i(D) - \sum_{C \subseteq \mcal C} \rho^c(C) \cdot v_i(C) -  \sum_{D \subseteq \mcal D} \rho^d(D) \cdot w_i(D)$$
        so the above set is $f^{-1}((0, \infty))$. By regularity condition (A1), $f$ is the linear combination of measurable functions and hence is measurable. As $(0, \infty)$ is a measurable set, $f^{-1}((0, \infty))$ is measurable, which finishes the proof of this claim.
    \end{proof}

    \noindent \textbf{Step Four: Closed Graph}

    Let 
    $$B(i, \hat{x}) = \{q \in \Delta (A): q \cdot U(i, \hat{x}) \geq q' \cdot U(i, \hat{x}) \text{ for all } q' \in \Delta (A)\}$$
    be the set of player $i$'s best responses to $\hat{x}$, where $q \cdot U(i, \hat{x}) = \sum_{a \in A} q(a) U_a(i, \hat{x})$ is the dot product between $q$ and $U(i, \hat{x})$.

    \begin{claim}
        Equip $\hat{A}$ with the weak topology. Then, for each $i$, the graph of $B(i,\cdot): \hat{A} \to \Delta A$ is closed in $\hat{A} \times \Delta (A)$.
    \end{claim}

    \begin{proof}
        Fix $i$. We will show that if $\hat{x}^n \to \hat{x}^0$ in $\hat{A}$ and $q^n \to q^0$ in $\Delta A$ such that $(\hat{x}^n, q^n)$ is in the graph of $B(i, \cdot)$ for each $n$, then $(\hat{x}^0, q^0)$ is in the graph of $B(i, \cdot)$ as well.

        Let $e_a$ denote the $a$-th basis vector in $\Delta (A)$, so 
        $$e_a(a') = \left\{\begin{matrix}
        1 & \text{ if } a' = a \\
        0 & \text{ otherwise.}\\
        \end{matrix}\right.$$ 
        By the definition of the weak topology, $\hat{x}^n \to \hat{x}^0$ implies that for any $a \in A$,
        $$\int_{\mcal I} \hat{x}^n_a(i) d\mu(i) = \int_{\mcal I} \hat{x}^n(i) \cdot e_a d\mu(i) \to \int_{\mcal I} \hat{x}^0(i) \cdot e_a d\mu(i) = \int_{\mcal I} \hat{x}^0_a(i) d\mu(i)$$
        as integrating against $e_a$ is a linear functional on $\hat{A}$. 

        Towards a contradiction, suppose $(\hat{x}^0, q^0)$ is \textit{not} in the graph of $B(i, \cdot)$, which means that for some $q^*$,
        $$q^0 \cdot U(i, \hat{x}^0) < q^* \cdot U(i, \hat{x}^0).$$
        Then,
        \begin{align*}
            & q^n \cdot U(i, \hat{x}^n) - q^* \cdot U(i, \hat{x}^n) \\
            = & \Big[q^n \cdot U(i, \hat{x}^n) - q^* \cdot U(i, \hat{x}^n)\Big] - \Big[q^0 \cdot U(i, \hat{x}^0) - q^* \cdot U(i, \hat{x}^0)\Big] + \Big[q^0 \cdot U(i, \hat{x}^0) - q^* \cdot U(i, \hat{x}^0)\Big] \\
            = & \Big[q^n \cdot U(i, \hat{x}^n) - q^0 \cdot U(i, \hat{x}^0)\Big] + q^* \cdot \Big[U(i, \hat{x}^0) - U(i, \hat{x}^n)\Big] + \Big[q^0 \cdot U(i, \hat{x}^0) - q^* \cdot U(i, \hat{x}^0)\Big]
        \end{align*}
        For the first term, 
        $$\Big[q^n \cdot U(i, \hat{x}^n) - q^0 \cdot U(i, \hat{x}^0)\Big] \leq \|q^n - q^0\|_{L^1} \cdot \max_a \left[\left|U_a(i, \hat{x}^n) - U_a(i, \hat{x}^0)\right|\right]$$
        which converges to zero as $\max_a \Big[\left|U_a(i, \hat{x}^n) - U_a(i, \hat{x}^0)\right|\Big]$ is bounded (there are only a finite number of bundles any player can get) and $q^n \to q^0$ implies $\|q^n - q^0\|_{L^1} \to 0$.

        For the second term, suppose $i$ receives a better bundle under $\hat{x}^0$ than $\hat{x}^n$ under some fixed tie-breaking rules $r^c, r^d$. Let $t_{\hat{x}}$ be the time where the bundle $i$ receives under $\hat{x}^0$ is no longer available, so 
        $$t_{\hat{x}} = \min_{c \in \hat{C}(i, \hat{x}(i)|\hat{x}, r^c)} i_c.$$ 
        It must be that $i$ chooses before $t_{\hat{x}^0}$ (but after $t_{\hat{x}^n}$). Then, there is an open interval of time between $i$ choosing and $t_{\hat{x}^0}$. As $t_{\hat{x}}$ varies continuously in $\left(\int_{\mcal I} \hat{x}_a(i) d\mu(i)\right)_{a \in A}$, $\hat{x}^n \to \hat{x}^0$ implies that $t_{\hat{x}^n} \to t_{\hat{x}^0}$. As such, the bundle $i$ receives under $\hat{x}^n$ will eventually be no worse (and potentially better) than the bundle they receive under $\hat{x}^0$. Since this holds for any $r^c, r^d$, as $n \to \infty$ the probability (with respect to nature drawing $r^c, r^d$) of $i$ receiving a better bundle under $\hat{x}^0$ than $\hat{x}^n$ tends to zero. As utilities are bounded, this in turn implies that $q^* \cdot \Big[U(i, \hat{x}^0) - U(i, \hat{x}^n)\Big]$ is non-positive for sufficiently large $n$.

        Finally, the third term is \textit{strictly} negative by assumption. Furthermore, the third term is independent of $n$, so for sufficiently large $n$, it must be that $q^n \cdot U(i, \hat{x}^n) - q^* \cdot U(i, \hat{x}^n) < 0$, which implies $q^n \cdot U(i, \hat{x}^n) < q^* \cdot U(i, \hat{x}^n)$. However, this contradicts $(q^n, \hat{x}^n)$ being in the graph of $B(i, \cdot)$.
    \end{proof}

    Claims 2 and 3 are Condition b and Claim 1 of \cite{schmeidler_1973_equilibrium}, which are sufficient to invoke their Theorem 1, guaranteeing the existence of a (potentially mixed-strategy) Nash equilibrium. Claim 1 establishes the additional hypothesis of \cite{schmeidler_1973_equilibrium}'s Theorem 2, which then guarantees the existence of a pure-strategy Nash equilibrium. This completes the proof.
\end{proof}

\section*{Appendix B: Omitted Proofs}
\makeatletter\def\@currentlabel{Appendix B}\makeatother
\label{Appendix B}

\subsubsection*{PROOF OF PROPOSITION \ref{preferences_seperation}}

\begin{proof}
    Let $\hat{s}(i)$ be equilibrium signals. Then, student $i$ reports $s$ to maximize
    $$\E{v_i\Big(\hat{C}(i, s| \hat{s}, r^c) \Big) + w_i\Big( \hat{D}(i, s| \hat{s}, r^d) \Big)}$$
    where the expectation is taken with respect to $r^c, r^d \sim R$. Suppose student $i$ cares more about the same courses than student $j$. For each fixed $r^c, r^d$, let ${r^c}', {r^d}'$ be the tie-breaking rules that permute $i$ and $j$'s tie-break priority (so $r^c(i) = {r^c}'(j)$). By symmetry, ${r^c}', {r^d}' \sim R$ as well. Then, the courses students $i$ and $j$ receive are identical up to tie-breaking as they have the same ordering over bundles of courses. As such,
    \begin{align*}
        & \left[v_i\Big(\hat{C}(i, s| \hat{s}, r^c) \Big) + w_i\Big( \hat{D}(i, s| \hat{s}, r^d) \Big)\right] - \left[v_j\Big(\hat{C}(j, s| \hat{s}, {r^c}') \Big) + w_j\Big( \hat{D}(j, s| \hat{s}, {r^d}') \Big)\right] \\
        = & v_i\Big(\hat{C}(i, s| \hat{s}, r^c) \Big) - v_j\Big(\hat{C}(j, s| \hat{s}, {r^c}')\Big)
    \end{align*}
    is increasing in $s$ since reporting a higher signal leads to a better course and $v_i - v_j$ is increasing by condition (2) of definition one. Taking expectations with respect to $r^c, {r^c}' \sim R$ gives that a student's expected utility function has increasing differences in the signal they report and caring more about the same courses. Thus, the monotone selection theorem of \cite{milgrom_1994_monotone} gives that a student who cares more about the same courses will report a higher signal.
\end{proof}

\subsubsection*{PROOF OF PROPOSITION \ref{competition_seperation}}

\begin{proof}
    The gain from reporting $s'$ compared to $\hat{s}(i)$ is
    $$\E{v_i\Big(\hat{C}(i, s'| \hat{s}, r^c) \Big) + w_i\Big( \hat{D}(i, s'| \hat{s}, r^d) \Big)} - \E{v_i\Big(\hat{C}(i, \hat{s}(i)| \hat{s}, r^c) \Big) + w_i\Big( \hat{D}(i, \hat{s}(i)| \hat{s}, r^d) \Big)}.$$
    As $i$ receives the same distribution of courses when reporting $s'$ as $\hat{s}(i)$, the above is equal to 
    $$\E{w_i\Big( \hat{D}(i, s'| \hat{s}, r^d) \Big)} - \E{w_i\Big( \hat{D}(i, \hat{s}(i)| \hat{s}, r^d) \Big)}.$$
    For each $r^d$, student $i$ picks dorms earlier when reporting $s'$ than $\hat{s}(i)$, so $w_i\Big( \hat{D}(i, s'| \hat{s}, r^d) \Big) - w_i\Big( \hat{D}(i, \hat{s}(i)| \hat{s}, r^d) \geq 0$. Integrating with respect to $r^d$ gives that there is a non-negative gain to reporting $s'$ instead of $\hat{s}(i).$
\end{proof}

\subsubsection*{PROOF OF PROPOSITION \ref{seperation_no_swap}}

\begin{proof}
    Suppose student $i$ cares more about the same courses than student $j$ and reports a different signal. By Proposition \ref{preferences_seperation}, student $i$ must have reported a higher signal than student $j$. Fix any arbitrary tie-breaking rules $r^c, r^d$ and let
    $$C_i = \hat{C}(i, s(i)|\hat{s}, r^c), D_i = \hat{D}(i, \hat{s}(i)|\hat{s}, r^d), C_j = \hat{C}(j, s(j)|\hat{s}, r^c), D_j = \hat{D}(j, \hat{s}(j)|\hat{s}, r^d).$$
    Then, it suffice to show that 
    $$u_i(C_i) + w_i(D_i) + v_j(C_j) + w_j(D_j) \geq v_i(C_j) + w_j(D_j) + v_j(C_i) + w_j(D_i)$$
    as then, it cannot be that both $i$ and $j$ are better off swapping. We have that
    \begin{align*}
        &[v_i(C_i) + w_i(D_i) + v_j(C_j) + w_j(D_j)] - [v_i(C_j) + w_i(D_j) + v_j(C_i) + w_j(D_i)] \\
        =& [v_i(C_i) - v_j(C_i)] - [v_i(C_j) - v_j(C_j)] \geq 0
    \end{align*}
    where the equality holds as $w_i(D_i) = w_j(D_i); w_i(D_j) = w_j(D_j)$ as $i$ caring more about the same courses than $j$ imply their utility over dorms are identical and the inequality holds as the difference in utility $i$ and $j$ get from courses is single-crossing and $C_i \succ C_j$ in the students' common ordering over courses since $\hat{s}(i) > \hat{s}(j)$ implies $i$ chooses before $j$ regardless of tie-break order.
\end{proof}

\subsubsection*{PROOF OF PROPOSITION \ref{FOSD_envy_free}}

\begin{proof}
    We will first prove the following Lemma. 

    \begin{lemma}\label{lemma1}
        Suppose re-labelings $\psi^C, {\psi^C}'$ are equal almost everywhere. If $\psi^C(i) = {\psi^C}'(j)$ then the set of courses available to $i$ when it is their turn to choose under $\psi^C$ is equal to the set of courses available to $j$ when it is their turn to choose under ${\psi^C}'$.
    \end{lemma}

    \begin{proof}
        It suffice to show that run-out times $i_c$ are unchanged between $\psi^C, {\psi^C}'$ for all courses.  

        From the proof of Proposition \ref{well_def} that given (potentially empty) $i_1,...,i_n; c_1,...,c_n$ we inductively have that
        $$i_{n+1} = \sup \left\{i: \min_{c \in \mcal C \setminus \{c_1,...,c_n\}} \left(q(c) - e_{i_n}(c) - \nu\Big(\{i_{c_n} \leq j < i: c \in C_j(\mcal C \setminus \{c_1,...,c_n\})\Big) \right) > 0\right\}$$
        where student identities are post-relabeling. As $\psi^C = {\psi^C}'$ almost everywhere, $\nu = \mu \circ (\psi^C)^{-1}$ is equal to $\nu' = \mu \circ ({\psi^C}')^{-1}$. As such,
        $$\nu\Big(\{i_{c_n} \leq j < i: c \in C_j(\mcal C \setminus \{c_1,...,c_n\})\Big) = \nu'\Big(\{i_{c_n} \leq j < i: c \in C_j(\mcal C \setminus \{c_1,...,c_n\})\Big)$$
        so all course run-out times are identical under $\psi^C$ and ${\psi^C}'$. Thus, all students with the same re-labeling under $\psi^C$ and ${\psi^C}'$ have the same set of available courses when it is their turn to choose.
    \end{proof}
    
    For any possible tie-breaking permutations of the unit interval, $r^c, r^d$ differ from $r^c \circ \sigma_{ij}, r^d \circ \sigma_{ij}$ on a set of zero measure. Then, letting $\psi^C, \psi^D$ be the re-orderings induced by $r^c, r^d$ and ${\psi^C}', {\psi^D}'$ be re-orderings induced by $r^c, r^d$, we have that $\psi^C = {\psi^C}', \psi^D = {\psi^D}'$ almost everywhere. By construction, $\psi^C(i) = {\psi^C}'(j)$ and $\psi^D(i) = {\psi^D}'(j)$. Lemma \ref{lemma1} then gives that sender $j$ has the same options under $r^c, r^d$ as sender $i$ has under $r^c \circ \sigma_{ij}, r^d \circ \sigma_{ij}$. As such,
    \begin{align*}
        \E{u_i(\hat{C}(i, \hat{s}(i)|\hat{s}, r^c), \hat{D}(i, \hat{s}(i)|\hat{s}, r^d))} &= \E{u_i(\hat{C}(i, \hat{s}(i)|\hat{s}, r^c \circ \sigma_{ij}), \hat{D}(i, \hat{s}(i)|\hat{s} \circ \sigma_{ij}, r^d))} \\
        & \geq \E{u_i(\hat{C}(j, \hat{s}(j)|\hat{s}, r^c), \hat{D}(j, \hat{s}(j)|\hat{s}, r^d))}
    \end{align*}
    where the equality follows from (A2) and the inequality comes from the fact that student $i$ could have always mimicked student $j$'s behavior but chose not to combined with the above discussion.
\end{proof}

\subsubsection*{PROOF OF THEOREM \ref{deterministic_pareto}}

\begin{proof}
    Let $\hat{s}$ be equilibrium signals and $m$ be the induced equilibrium allocation. Without loss of generality, let $d_k$ be the $k$th dorm to reach capacity. Similarly, let $s_k$ be the smallest signal for which dorm $d_k$ has not yet reached capacity. Let $t: \mcal C \cup \mcal D \to [0, 1]$ be a function mapping courses and dorms to the time they are at capacity at the original equilibrium. For each $s \in \mcal S$, let 
    $$t_s = \mu(\{i: \hat{s}(i) \geq s'\})$$
    be the mass of students who reveal a signal above $s$. As such, students that reveal $\hat{s}(i) = s$ select courses between time $t_s$ and $t_{s+1}$ and dorms between time $1-t_{s+1}$ and $1-t_s$ depending on randomization. The following Lemma then establishes that any student who selects signal $s_k$ has the option to choose dorm $d_k$. 

    \begin{lemma}\label{deterministic_times}
        If $m$ is deterministic then run-out times are deterministic and are a subset of signal cutoff times: $\{t(c), t(d): c, d \in \mcal C \cup \mcal D\} \subseteq \{t_s: s \in \mcal S\}$.
    \end{lemma}

    \begin{proof}
        Towards a contradiction, suppose there exists some course $c$ (the case where some dorm violates the statement is symmetric) such that $t(c) \notin \{t_s : s \in \mcal S\}$ with positive probability. Then, there exists some $s$ such that $t_s < t(c) < t_{s+1}$ so the set of students $i$ with $\hat{s}(i) = s$ and $m_c(i) = c$ is of positive measure. Each student in this set has a course draw time uniformly distributed over $[t_s, t_{s+1}]$ so the probability that a positive mass of students all have course draw time in $[t_s, t(c)]$ is zero. 
        
        Next, if $t(c)$ is not deterministic and the support of $t(c)$ cannot be bound in some $[t_s, t_{s+1}]$, then there must exist some $s$ such that $t(c) < t_s$ with positive probability and $t(c) \geq t_s$ with positive probability. Then, some positive measure of students revealing $s$ receive $c$ with positive probability but not probability one, a contradiction.
    \end{proof}
    
    Let $\mcal I_k = \{i: m_d'(i) = d_k\}$. Note that each student $i \in \mcal I_k$ has to reveal an equilibrium signal $\hat{s}(i)$ lower than $s_k$, as otherwise they would not have been able to select dorm $k$. We will show that if allocation $m'$ satisfies $m'(i) \succeq m(i)$ for all $i$, then $m' = m$. We will do so by inducting (in reverse order) on $k$ and showing that $m'(i) = m(i)$ for all $i$ in $\mcal I_k$. Fix $k \in \{1,...,n\}$ and suppose 
    $$m'(i) = m(i) \text{ for all } i \in \bigcup_{\ell > k} \mcal I_\ell.$$
    For the inductive step, we will show that $m'(i) = m(i)$ for all $i \in \mcal I_k$ (the base case for induction is when $k = n$). Fix $i \in \mcal I_k$. If under equilibrium play $\hat{s}$ student $i$ reveals $s_k$, potentially deviating from $\hat{s}(i)$, they can select any dorm $d_k$ (or lower) and any bundle of courses not yet filled with students from $\bigcup_{\ell > k} \mcal I_\ell$. Let $(C', d')$ be the bundle $i$ gets from this deviation. By the inductive hypothesis, $m = m'$ on $\in \bigcup_{\ell > k} \mcal I_\ell$ so $(C', d') \succeq m'(i)$. For this deviation to not have been profitable, it must be that $m(i) \succeq (C', d')$. Putting everything together, we get that
    $$m(i) \succeq (C', d') \succeq m'(i) \succeq m(i).$$
    As such, $m(i) \sim m'(i)$. Thus, there cannot be any Pareto improvements over $m$ as if each student is weakly better off, no student can be strictly better off.

    Finally, if there were any student $i$ that envied student $j$, then $i$ can always replicate $j$'s behavior to get $j$'s bundle.
\end{proof}

\subsubsection*{PROOF OF PROPOSITION \ref{signal_expansion}}
\begin{proof}
    As signals only matter ordinally, let $\mcal S = \{1,2,...,k\}$ and $\mcal S' = \{1,2,...,K\}$ for $K > k$. Then, $\hat{s}$ is still an equilibrium when the signal space is $\mcal S'$.\footnote{In the language of the original Proposition, it may not be the case that $\hat{s} = \hat{s}'$ since signals were re-labeled. It could be the case that $\mcal S$ does not consist of the $k$ smallest signals of $\mcal S'$.} For each student $i$, deviating from $\hat{s}(i)$ to $s \in \{1,2,...,k\}$ cannot be profitable as $\hat{s}$ was an equilibrium when the signal space is $\mcal S$ and such deviations were previously possible. Finally, deviating from $\hat{s}(i)$ to $s \in \{k+1,k+2,...,K\}$ is weakly dominated by revealing $s = k$.
\end{proof}

\subsubsection*{PROOF OF PROPOSITION \ref{ex_ante_improvement_1}}

\begin{proof}
    Let $\hat{s}$ be an equilibrium of the paired serial dictatorship mechanism. Since all students have the same utility function over courses $v$ and same utility function over dorms $w$, each student chooses the same set of courses and the same dorm when the set of available courses or dorms is the same. As such, a student's allocation is determined solely by \textit{when} they choose regardless of \textit{which} students have already chosen.

    Then, each student $i$ receives the same (marginal) distribution over draw times in each market under paired serial dictatorship by matching their signal with the overall distribution of signals and playing the mixed strategy
    $$\hat{s}'(i) = \delta_{s} \text{ with probability } \mu(\{j \in \mcal I: \hat{s}(j) = s\}) $$
    as under independent RSD. As $\hat{s}(i)$ is a best response to $\hat{s}$, it must be that $i$ is weakly better off playing $\hat{s}$ than $s'$.
\end{proof}

\subsubsection*{PROOF OF THEOREM \ref{ex_ante_efficienct_1}}
\begin{proof}
    We will first use optimal transport methods to characterize the efficient allocation given some fixed signal space $\mcal S$ and then verify that it is an equilibrium of paired serial dictatorship.

    Let $\Lambda$ denote the set of all possible $\lambda$, $F$ denote the distribution of $\lambda$ in the population, $q_c = q|_{\mcal C}$ denote the quantity of courses, and $q_d = q|_{\mcal D}$ denote the quantity of courses. As students have a common ordering over courses and dorms, let 
    $$Q_c(c) = \int_{c': v(c') \leq v(c)} q_c(c') dc'; Q_d(d) = \int_{d': w(d') \leq w(d)} q_d(d') dd'$$
    be cumulative distributions over courses or dorms. Then, the ex-ante optimal allocation when students can only differentiate themselves via signals in $\mcal S$ solves the following optimal transport problem:
    \begin{align}\label{OT1}
    \begin{split}
        \max_{A \in \Delta (\Lambda, \mcal S, \mcal C, \mcal D)} & \left\{\int_{\mcal S \times \Lambda \times \mcal C \times \mcal D} \left[\frac{\lambda}{1+\lambda}v(c) + \frac{1}{1+\lambda}w(d)\right] dX(s, \lambda, c, d)\right\} \\
        \text{s.t.   } & (X_{\mcal S, \mcal C, \mcal D} | s) \independent (X_{\Lambda, \mcal S} | s) \\
        & X_{\mcal C} = Q_c, X_{\mcal D} = Q_d, X_\Lambda = F.
    \end{split}
    \end{align}
    A coupling $X$ over student parameters, signals, courses, and dorms pins down an assignment. The objective is to maximize total expected utility over all possible assignments. The first constraint specifies that course and dorm assignments can only be coupled with student parameters through their signals. The remaining constraints specify that the distribution of courses, dorms, and student parameters match the underlying environment. 

    By conditional independence and the additive separability, we can reformulate (\ref{OT1}) by splitting the joint distribution $A$ into three couplings between students and signals, signals and courses, and signals and dorms:
    \begin{align}\label{OT2}
    \begin{split}
        \max_{\substack{X \in \Delta(\Lambda, \mcal S), \\ Y \in \Delta(\mcal S, \mcal C), \\ Z \in \Delta(\mcal S, \mcal D)}} & \left\{\int_{\Lambda \times \mcal C} \left[\frac{\lambda}{1+\lambda} v(c)\right] d(X \circ Y)(\lambda, c) + \int_{\Lambda \times \mcal D} \left[\frac{1}{1+\lambda} w(d)\right] d(X \circ Z)(\lambda, d)\right\} \\
        \text{s.t.   } & X_\Lambda = F, Y_{\mcal C} = Q_c, Z_{\mcal D} = Q_D \\
        & X_{\mcal S} = Y_{\mcal S} = Z_{\mcal S}.
    \end{split}
    \end{align}
    The first constraint specifies that the distribution of courses, dorms, and student parameters match the underlying environment while the second constraint ensures that students and courses or dorms can in fact be coupled through signals. This assumption ensures that 
    $$(X \circ Y)(\lambda, c) = \int_{\mcal S} X(\lambda, s) Y(s, c) ds, (X \circ Z)(\lambda, d) = \int_{\mcal S} X(\lambda, s) Z(s, d) ds$$
    are probability measures on $\Lambda \times \mcal C, \Lambda \times \mcal D$ themselves. 
    
    We now analyze solutions to (\ref{OT2}). For any $X$ that satisfies the constraints of (\ref{OT2}), the problem becomes solving
    \begin{align}\label{OT3a}
    \begin{split}
        \max_{Y \in \Delta(\mcal S, \mcal C)} & \left\{\int_{\mcal S \times \mcal C} \left[\int_\Lambda \frac{\lambda}{1+\lambda} dX(\lambda, s) \right]\cdot v(c) dY(s', c) \right\} \\
        \text{s.t.   } & Y_{\mcal C} = Q_c, Y_{\mcal S} = X_{\mcal S}.
    \end{split}
    \end{align}
    and
    \begin{align}\label{OT3b}
    \begin{split}
        \max_{Z \in \Delta(\mcal S, \mcal D)} & \left\{\int_{\mcal S \times \mcal D} \left[\int_\Lambda \frac{1}{1+\lambda} dX(\lambda, s) \right] \cdot w(d) dZ(s', d) \right\} \\
        \text{s.t.   } & Y_{\mcal D} = Q_d, Z_{\mcal S} = X_{\mcal S}.
    \end{split}
    \end{align}
    We will focus on (\ref{OT3a}). Order $\mcal S$ so 
    $$s \geq s' \text{ if and only if } \int_\Lambda \frac{\lambda}{1+\lambda} dX(\lambda, s) \geq \int_\Lambda \frac{\lambda}{1+\lambda} dX(\lambda, s')$$
    and order $\mcal C$ according to the usual order ($c \geq c'$ if and only if $v(c) \geq v(c')$). Then, the integrand 
    $$\left[\int_\Lambda \frac{\lambda}{1+\lambda} dX(\lambda, s) \right]\cdot v(c)$$
    is supermodular in $s, c$. As such, the optimal transport $Y$ is co-monotone between $\mcal S$ and $\mcal C$.\footnote{See \cite{filipposantambrogio_2015_optimal} or \cite{luo_2024_marginal}.} Similarly, the transport $Z$ that solves (\ref{OT3b}) is co-monotone between $\mcal S$ and $\mcal D$ when $\mcal S$ is ordered according to
    $$s \geq s' \text{ if and only if } \int_\Lambda \frac{1}{1+\lambda} dX(\lambda, s) \geq \int_\Lambda \frac{1}{1+\lambda} dX(\lambda, s').$$

    Next, for any $Y, Z$ that satisfies the constraints of (\ref{OT2}), the problem becomes 
    \begin{align}\label{OT4}
    \begin{split}
        \max_{X \in \Delta(\Lambda, \mcal S)} & \left\{\int_{\Lambda \times \mcal S}\frac{\lambda}{1+\lambda} \left[\int_{\mcal C} v(c) dY(s, c)\right] + \frac{1}{1+\lambda} \left[ \int_{\mcal D} w(d) dZ(s, z)\right] dX(\lambda, s)\right\} \\
        \text{s.t.   } & X_\Lambda = F\\
        & X_S = Y_S = Z_S.
    \end{split}
    \end{align}
    Now, order $\mcal S$ so $\int_{\mcal C} v(c) dY(s, c) - \int_{\mcal D} w(d) dZ(s, z)$ is increasing in $s$. Then, 
    $$\frac{\lambda}{1+\lambda} \left[\int_{\mcal C} v(c) dY(s, c)\right] + \frac{1}{1+\lambda} \left[ \int_{\mcal D} w(d) dZ(s, z)\right] $$
    is supermodular in $\Lambda, \mcal S$ (where the usual order is imposed on $\Lambda$) as
    \begin{align*}
        &\frac{\partial}{\partial \lambda}\left(\frac{\lambda}{1+\lambda} \left[\int_{\mcal C} v(c) dY(s, c)\right] + \frac{1}{1+\lambda} \left[ \int_{\mcal D} w(d) dZ(s, z)\right]\right) \\
        = &\frac{1}{(\lambda+1)^2} \left(\int_{\mcal C} v(c) dY(s, c) - \int_{\mcal D} w(d) dZ(s, z)\right)
    \end{align*}
    is increasing in $\mcal S$. As such, it suffice to look at solutions for which there exists an order on $\mcal S$ making $X$ co-monotone between $\Lambda, \mcal S$, $Y$ co-monotone between $\mcal S, \mcal C$, and $Z$ co-monotone between $\mcal S$ with the reverse order and $\mcal D$. Thus, $Y$ and $Z$ are pinned down by a joint distribution $X$. Suppose that $F$ has density $f$ and let $\mcal S = \{1,...,n\}$ so $X$ can be parameterized by $n-1$ threshold values of $\lambda$. Denote them $\lambda_1,...,\lambda_{n-1}$ and set $\lambda_0 = \inf \Lambda, \lambda_n = \sup \Lambda$. Then,
    $$X(\lambda, s) = \begin{cases}
        \frac{f(\lambda)}{(F(\lambda_s) - F(\lambda_{s-1}))^2}& \text{ if } \lambda \in (\lambda_{s-1}, \lambda_s) \\
        0 & \text{ else}.
    \end{cases}$$
    An example of one such $X$ with $\mcal S = \{1, 2, 3\}, \Lambda = [0,1], F = U(0,1), \lambda_0 = 0, \lambda_1 = 1/2, \lambda_2 = 2/3, \lambda_3 = 1$ is shown in Figure \ref{coarse_OT}:
    
    \begin{figure}[h]
        \centering
        \begin{tikzpicture}

        \draw[-] (0,0) -- (5,0) node[right] {$\Lambda$};
        \draw[-] (0,0) -- (0,5) node[above] {$\mcal S$};
        \draw[-] (0,5) -- (5,5) -- (5,0);
        
        \draw (0,0) -- (0,-0.2) node[below] {$0$};
        \draw (2.5,0) -- (2.5,-0.2) node[below] {$\frac{1}{2}$};
        \draw (3.333,0) -- (3.333,-0.2) node[below] {$\frac{2}{3}$};
        \draw (5,0) -- (5,-0.2) node[below] {$1$};
        \draw[red] (0,1.25) -- (-0.2,1.25) node[left] {$1$};
        \draw[blue] (0,2.9166) -- (-0.2,2.9166) node[left] {$2$};
        \draw[green] (0,4.1666) -- (-0.2,4.1666) node[left] {$3$};
        
        \fill[red!50,opacity=0.5] (0,0) -- (2.5,0) -- (2.5,2.5) -- (0,2.5) -- cycle;
        \fill[blue!50,opacity=0.5] (2.5,2.5) -- (3.333,2.5) -- (3.333,3.333) -- (2.5,3.333) -- cycle;
        \fill[green!50,opacity=0.5] (3.333,3.333) -- (5,3.333) -- (5,5) -- (3.333,5) -- cycle;
        
        \draw[dashed] (0,2.5) -- (5,2.5);
        \draw[dashed] (2.5,0) -- (2.5,5);
        \draw[dashed] (3.333,0) -- (3.333,5);
        \draw[dashed] (0,3.333) -- (5,3.333);
        \draw[dashed] (0, 0) -- (5,5);
        
        \draw[thick, red] (0,0) -- (2.5,0) -- (2.5,2.5) -- (0,2.5) -- cycle;
        \draw[thick, blue] (2.5,2.5) -- (3.333,2.5) -- (3.333,3.333) -- (2.5,3.333) -- cycle;
        \draw[thick, green] (3.333,3.333) -- (5,3.333) -- (5,5) -- (3.333,5) -- cycle;
        
        \draw[red,thick] (6,4.5) -- (6.5,4.5);
        \node[red, right] at (6.5,4.5) {\small mass $1/2$};
        \draw[blue,thick] (6,4) -- (6.5,4);
        \node[blue, right] at (6.5,4) {\small mass $1/6$};
        \draw[green,thick] (6,3.5) -- (6.5,3.5);
        \node[green, right] at (6.5,3.5) {\small mass $1/3$};
        
        \end{tikzpicture}
        \caption{Example of Coarse Transport}
        \label{coarse_OT}
    \end{figure}
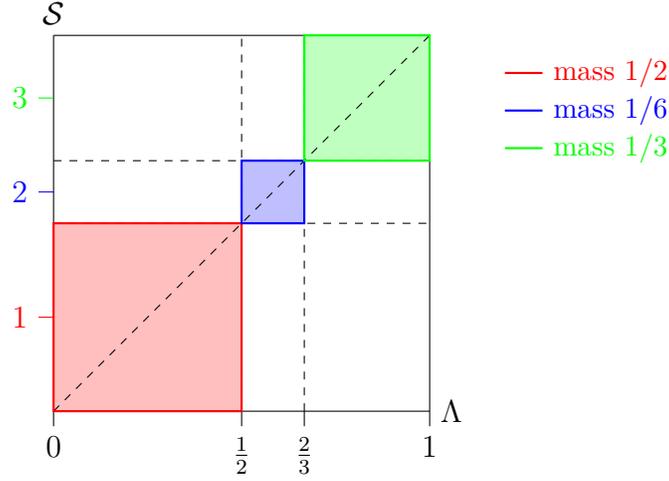
    
    Given an $X$ (induced by $\lambda_1,...,\lambda_{n-1}$), let $Y^*$ be the solution to (\ref{OT3a}) and let $Z^*$ be the solution to (\ref{OT3b}). With this formulation, we can write the original problem (\ref{OT1}) as:
    \begin{align}\label{OT5}
    \begin{split}
        \max_{\lambda_1,...,\lambda_{n-1}} & \left\{ \sum_{i=1}^n \left( \int_{\lambda_{i-1}}^{\lambda_i} \frac{\lambda}{1+\lambda} \E[Y^*]{v(c)|s_i} + \frac{1}{1+\lambda} \E[Z^*]{w(d)|s_i} dF(\lambda) \right) \right\}.
    \end{split}
    \end{align}
    Next, we consider the first order conditions of (\ref{OT5}). By the envelope theorem, we can hold $Y^*, Z^*$ constant when differentiating with respect to $\lambda_i$. Increasing $\lambda_i$ increases the mass associated with signal $i$ but decreases the mass associated with signal $i+1$, so by the Fundamental Theorem of Calculus, letting $O(\lambda_1,...,\lambda_{n-1})$ be the objective in (\ref{OT5}), we have that the first order condition with respect to $\lambda_i$ is
    \begin{align*}
    \begin{split}
        0 = \frac{\partial}{\partial \lambda_i} O(\lambda_1,...,\lambda_{n-1}) =& \left(\frac{\lambda_i}{1+\lambda_i} \E[Y^*]{v(c)|s_i} + \frac{1}{1+\lambda_i} \E[Z^*]{w(d)|s_i}\right)f(\lambda_i) \\
        &- \left(\frac{\lambda_i}{1+\lambda_i} \E[Y^*]{v(c)|s_{i+1}} + \frac{1}{1+\lambda_i} \E[Z^*]{w(d)|s_{i+1}}\right)f(\lambda_i)
    \end{split}
    \end{align*}
    which implies that 
    $$\frac{\lambda_i}{1+\lambda_i} \E[Y^*]{v(c)|s_i} + \frac{1}{1+\lambda_i} \E[Z^*]{w(d)|s_i} = \frac{\lambda_i}{1+\lambda_i} \E[Y^*]{v(c)|s_{i+1}} + \frac{1}{1+\lambda_i} \E[Z^*]{w(d)|s_{i+1}}.$$
    As such, no student on the boundary would prefer to deviate to revealing a higher (and by symmetry, lower) signal. 

    We will now use the above first order condition to show how the optimum can be sustained in equilibrium. As preferences over courses or dorms are common, students that reveal higher signals receive a better course and a worse dorm. As such, any coupling between signals and courses or dorms induced by an equilibrium of paired serial dictatorship is co-monotone in the same way as identified above. By Proposition \ref{preferences_seperation}, an equilibrium coupling between $\Lambda$ and $\mcal S$ will also be co-monotone. By independent tie-breaking within signals, each student only cares about the expected course or dorm quality at each signal.

    Finally, no students would want to deviate from the optimum. Towards a contradiction, suppose a student with $\lambda \in (\lambda_{i-1}, \lambda_{i})$ would strictly benefit from reporting signal $j > i$ instead of signal $i$ (downwards deviations can be handled similarly). By Proposition \ref{preferences_seperation}, student payoffs have increasing differences in $\lambda$ and signal, so any student with $\lambda' \geq \lambda$ would also gain at least as much as that student from deviating. This implies that a student with $\lambda' > \lambda_{j-1} \geq \lambda_i > \lambda$ would also strictly benefit from deviating between revealing $j-1$ and $j$. However, the first order condition for $\lambda_j$ at the optimum gives that this student must be indifferent between revealing signal $j-1$ and signal $j$, a contradiction.
\end{proof}

\end{document}